\documentclass[12pt]{article}
\usepackage{amsmath,amssymb,amsthm}
\usepackage{booktabs}
\usepackage{multirow}
\usepackage{siunitx}
\usepackage{graphicx}
\usepackage{geometry}
\usepackage{algorithm}
\usepackage{algorithmicx}
\usepackage{comment}
\usepackage{subcaption}
\usepackage{algpseudocode}
\usepackage{pdfpages}
\usepackage{bm}
\usepackage{natbib}
\usepackage{hyperref}
\usepackage{adjustbox}
\usepackage{float}
\usepackage{booktabs}
\usepackage{longtable}
\usepackage{pdflscape}
\usepackage{epstopdf}
\usepackage{comment}
\usepackage{threeparttable}

\title{Interval-valued circular time series: A cylindrical
score-driven approach with an application to wind direction}

\author{
Thasmika Mohan$^{1}$, Najmeh Nakhaei Rad$^{1}\footnote{Correspoding Author: najmeh.nakhaeirad@up.ac.za}$, 
Sphiwe Skhosana$^{1}$ \\
Yaser Samadi $^{2}$\\[4pt]
\small $^{1}$Department of Statistics, University of Pretoria, Pretoria 0002, South Africa. \\
\small $^{2}$School of Mathematical and Statistical Sciences, Southern Illinois University, Carbondale, IL, USA.
}

\date{}
\begin{document}
\maketitle 
\begin{abstract} 
Interval-valued circular observations arise naturally in a wide range of environmental applications, particularly in wind direction monitoring, where directional measurements are observed by an interval of recorded directions over a period. An existing dynamic approach for interval-valued circular time series models the interval boundaries directly on the torus. While such an approach preserves the circular nature of the endpoints, an alternative approach is to consider the centre-range representation of circular intervals. This paper proposes a cylindrical score-driven framework for interval-valued circular time series by representing each interval through a circular centre and a bounded linear width. Dynamic conditional score recursions are developed for the centre and range processes, allowing circular-linear dependence to be captured jointly within a likelihood-based framework. To develop this model, various cylindrical distributions are introduced to accommodate bounded range values. The estimators are investigated through simulations, and the practical usefulness of the framework is illustrated using interval-valued wind direction data from South African and Brazilian wind farms. The proposed methodology provides an alternative approach for modelling the interval-valued circular processes using the directional location and dispersion in interval-valued circular observations. 
\end{abstract}

\section{Introduction}

Circular data, recorded as angles or directions on the unit circle, arise in many disciplines,
with wind direction among the most widely studied examples \citep{mardia2009directional,
pewsey2021recent}. Their defining feature is periodicity. Circular
variables also play an important role in analysing and forecasting the consequences of climate change, and wind direction in particular is a key input for wind-farm placement, power production and disaster management, such as for predicting the
spread of wildfires \citep{rad2022enhancing, ridder2025power, harvey2024modelling, nakhaei2026predicting}.

On wind farms, sensors routinely record the wind direction over short windows as intervals; for example, wind direction is recorded every 10 minutes using the minimum and maximum directions observed (anticlockwise from 0 to $2\pi$). Modelling the lower and the upper bounds separately disregards information, as
successive intervals are dependent. For example, the wind direction at one time point is rarely independent of its direction at the moment before. Interval-valued circular time series, therefore,
require models that respect the circular geometry, retain the interval structure and capture
temporal dependence.

Two models have recently been proposed for circular interval-valued data: (i) the circular interval-valued autoregressive (CINAR) model \citep{mohan2026cinar} uses a link function to map the interval endpoints to the real line \citep{fisher1992regression} and considers internal distributions for observations within intervals. This model does not perform any aggregation and uses the full information within the intervals. However, because of the limitations of this transformation, the model is mainly suitable for intervals concentrated within a small arc away from the circular cut; (ii) the interval-valued dynamic score-driven (IN-DCS) model \citep{mohan2026interval} instead represents the two endpoints jointly as a pair of circular observations on the torus. It models the endpoints directly, without a link function, and is fully rotationally invariant. It can therefore accommodate both small and large arcs, including intervals near or crossing the circular cut. The IN-DCS models the lower and upper boundary processes jointly called min--max aggregation on the real line \citep{neto2008centre}. Alternatively, an interval can be expressed by its centre on the circle and the arc length between the lower and upper boundaries as a range. On the real line, this aggregation of interval observations is known as a centre--range aggregation \citep{neto2008centre, maia2008forecasting}, however this has not been developed for circular interval-valued observations.

The centre of an arc is a circular observation, and its width
is a bounded linear quantity. Each observation, therefore, lies on a cylinder rather than a torus, as in the IN-DCS model  \citep{mohan2026cinar}. Dynamic models for
cylindrical time series data have been developed by \cite{harvey2023regime}, and these methods are suitable for modelling wind direction and speed, where the linear component is
unbounded. However, the range of a circular interval-valued observation is, contrastingly, bounded by the geometry of the circle.

This paper proposes a cylindrical score-driven framework for interval-valued circular time series using a centre--range approach. The contributions are as follows.
\begin{itemize}
  \item A novel centre--range representation of interval-valued circular observations modelled on the bounded cylinder $S^1 \times (0,1)$ is proposed.
  \item Three new cylindrical distributions are introduced for the joint distribution of centre and range, namely the Kumaraswamy--von Mises (KvM), beta--von Mises (BvM) and truncated Weibull--von Mises (TWvM) models.
 \item Three novel cylindrical score-driven time series models for interval-valued circular data are proposed: the centre--range truncated Weibull--von Mises (CR-TWvM), the centre--range beta-- von-Mises (CR-BvM) and the centre--range Kumaraswamy--von-Mises (CR-KvM).
\end{itemize}

The remainder of the paper is organised as follows. Section~\ref{sec:background} reviews
score-driven models for circular time series.
Section~\ref{sec:methodology} introduces the three proposed time series models and their estimation procedures;
in Section~\ref{sec:simulation} a simulation study is provided followed by a synthetic data analysis in Section \ref{sec:synthetic}. The models are fitted to two real-world datasets in South Africa and Brazil in Section~\ref{sec:application}, and lastly a conclusion is provided in Section \ref{sec:conclusion}.

\section{Background on score-driven time series models}\label{sec:background}
The score-driven, or dynamic conditional score (DCS) approach was introduced by \cite{harvey2013} and \cite{creal2013} on the real line. Later,  \cite{harvey2024modelling} developed this approach for circular time series data where the observations, $\theta_t \in [-\pi, \pi)$, follow a von Mises distribution,
\begin{equation*}
  f(\theta_t;\mu,\upsilon) = \frac{1}{2\pi I_0(\upsilon)}\exp\{\upsilon\cos(\theta_t-\mu)\},
\end{equation*}
where $I_0(\cdot)$ is the modified Bessel function of order zero, $\mu$ is the mean direction and
$\upsilon \ge 0$ is the concentration parameter. The location evolves as
\begin{equation*}
  \theta_t = \mu + \mu_{t|t-1} + \varepsilon_t, \qquad
  \mu_{t+1|t} = \phi\,\mu_{t|t-1} + \kappa\,u_t, \qquad
  u_t = \sin(\theta_t - \mu - \mu_{t|t-1}),
\end{equation*}
where $\kappa$ is the gain, $\phi$ is the
persistence parameter and $u_t = \sin(y_t - \mu - \mu_{t|t-1}) \in [-1, 1]$ is the forcing
variable. For more information, see \cite{harvey2024modelling}. 

Approaches to modelling cylindrical time series include hidden Markov models and score-driven models. Hidden Markov models describe transitions between latent regimes, each characterised by a different joint distribution of the circular and linear components \citep{lagona2015hidden}. Some formulations assume that these components are independent conditional on the regime. However, this assumption can be restrictive for variables such as wind speed and direction. \citet{mastrantonio2015bayesian} relax this assumption using a circular--linear general projected normal distribution. Within the score-driven framework, \citet{harvey2023regime} developed dynamic models based on the Weibull--sine-skewed--von Mises cylindrical distribution (Abe-Ley distribution) \citep{abe2017tractable}. The circular location and linear scale are updated using conditional likelihood scores, with extensions allowing time-varying concentration and regime switching. They illustrate the approach using hourly wind direction and speed data from Galicia, Spain.

Most recently, \citet{mohan2026interval} proposed the first-ever score-driven approach for modelling interval-valued circular time series data, which
models each interval through its lower and upper bounds by considering the toroidal Sine distribution
of \citet{singh2002probabilistic},
\begin{equation}
  f\!\left(\varepsilon^L_t, \varepsilon^U_t\right) = C(\upsilon_L, \upsilon_U, \lambda)^{-1}
  \exp\!\left\{\upsilon_L \cos \varepsilon^L_t + \upsilon_U \cos \varepsilon^U_t + \lambda \sin \varepsilon^L_t \sin \varepsilon^U_t\right\},
  \label{eq:bvsine}
\end{equation}
where $C(\cdot)$ is the normalising constant and $\lambda$ governs the dependence between the bounds,  $\lambda^2 < \upsilon_L\upsilon_U$ ensuring unimodality and $\lambda = 0$ gives the independent
model.  To consider an alternative framework for circular interval-valued time series, the next section discusses a cylindrical score-driven method using centre-range aggregation. 

\section{Methodology}
\label{sec:methodology}

In this section, a score-driven model for interval-valued circular time series is proposed. As an alternative to modelling the lower and upper boundary processes of an interval, each interval is
represented by its circular centre and its range such that an observation lies on the
bounded cylinder $\mathbb{S}^1\times(0,1)$. Let
\[
\Theta_t=[\theta_{t}^L,\theta_{t}^U], \qquad t=1,\ldots,T,
\]
denote a sequence of interval-valued circular observations, where $\theta_{t}^L$ and $\theta_{t}^U \in [-\pi, \pi)$ are
the lower and upper bounds of the interval recorded at time $t$. The circular centre ($C_t$) is
\begin{equation}
  C_t =
  \begin{cases}
    \mathrm{mod}_{2\pi}\!\left(\dfrac{\theta_{t}^L + \theta_{t}^U}{2} + \pi\right) - \pi, & \theta_{t}^U \ge \theta_{t}^L,\\[2ex]
    \mathrm{mod}_{2\pi}\!\left(\dfrac{\theta_{t}^L + \theta_{t}^U}{2}\right) - \pi,       & \theta_{t}^U <   \theta_{t}^L.
  \end{cases}
\end{equation}
The range of the circular interval is given as
\begin{equation}
  Q_t = \frac{\mathrm{mod}_{2\pi}\big(\theta_{t}^U - \theta_{t}^L\big)}{2\pi} ,
\end{equation}
where $Q_t \in (0, 1)$.
The lower and upper bounds of the interval can therefore be written as

\begin{equation*}
  \theta_{t}^L = \mathrm{mod}_{2\pi}\big(C_t - \pi Q_t + \pi\big) - \pi, \qquad
  \theta_{t}^U = \mathrm{mod}_{2\pi}\big(C_t + \pi Q_t + \pi\big) - \pi .
\end{equation*}

\subsection{Conditional cylindrical distributions for centre--range observations}

Let $\mathcal F_{t-1}$ denote the information available up to time
$t-1$. We model the centre--range observations $(C_t,Q_t)$ by specifying
the distribution of the range and the conditional distribution of the
centre given the range. Following the circular modelling framework of
\citet{harvey2024modelling}, we write the observation equation as
\[
C_t =\mu_C+\mu_{t|t-1}+\varepsilon_t,
\]
where $\mu_C$ is a baseline direction and $\mu_{t|t-1}$ is a
time-varying location component determined by past observations.
Conditional on $Q_t=q_t$ and $\mathcal F_{t-1}$, the error
$\varepsilon_t$ follows a von Mises distribution with location zero
and concentration $\upsilon_t(q_t)$. The conditional density of the centre is therefore
\begin{equation}
f_{C\mid Q,\mathcal F_{t-1}}(c_t\mid q_t)
=
\frac{
\exp\!\left\{
\upsilon_t(q_t)
\cos\!\left(c_t-\mu_C-\mu_{t|t-1}\right)
\right\}
}{
2\pi I_0\!\left(\upsilon_t(q_t)\right)
},
\qquad -\pi\leq c_t<\pi,
\label{eq:vm}
\end{equation}
where $I_0$ is the modified Bessel function of the first kind and order zero.

We allow the concentration of the centre to depend on the range
through
\begin{equation}
\upsilon_t(q_t)=\exp(\zeta-\delta q_t),
\qquad \zeta,\delta\in\mathbb R.
\label{eq:dependence}
\end{equation}
 If $\delta=0$, the centre
and range are conditionally independent given $\mathcal F_{t-1}$, and $\zeta$ is the log-concentration parameter. This construction follows the cylindrical
modelling approach of \citet{harvey2023regime}.

We consider three distributions for the normalised range
$Q_t\in(0,1)$. The first is a truncated Weibull distribution,
which adapts the Weibull-based approach of
\citet{harvey2023regime} to the bounded support of the range.
The second is the reparameterisation of the beta distribution \citep{ferrari2004beta}.
The third is the Kumaraswamy distribution
\citep{kumaraswamy1980}, which has a simple density and closed-form
distribution and quantile functions \citep{jones2009}.

The choice of range distribution affects the score-driven updates
because the three distributions produce different scores.
We compare the three specifications
using model selection criteria and predictive performance, since no
distribution is uniformly preferable.

For each model, the joint conditional density is
\begin{equation}
\begin{aligned}
f(c_t,q_t\mid\mathcal F_{t-1})
&=
f_Q(q_t\mid\mathcal F_{t-1})
f_{C\mid Q,\mathcal F_{t-1}}(c_t\mid q_t)\\
&=
f_Q(q_t\mid\mathcal F_{t-1})
\frac{
\exp\!\left\{
\upsilon_t(q_t)
\cos\!\left(c_t-\mu_C-\mu_{t|t-1}\right)
\right\}
}{
2\pi I_0\!\left(\upsilon_t(q_t)\right)
}.
\end{aligned}
\label{eq:cr-density}
\end{equation}
The log-likelihood function
at time $t$ is
\begin{equation}
\ell_t
={}
\log f_Q(q_t)
+\upsilon_t(q_t)
\cos\!\left(c_t-\mu_C-\mu_{t|t-1}\right)
-\log(2\pi)
-\log I_0\!\left(\upsilon_t(q_t)\right).
\label{eq:cr-loglik}
\end{equation}

The models differ in their range distribution, whose time-varying
parameter is linked to the filter $\eta_{t|t-1}$. For each model,
we give the density, the range score
$s_{Q,t}=\frac{\partial\ell_t}{\partial\eta_{t|t-1}}$,
and the conditional mean used for one-step-ahead range prediction.
Since the conditional centre density does not depend on
$\eta_{t|t-1}$ in this specification, the range score is obtained
by differentiating $\log f_Q(q_t)$.

\subsubsection{Truncated Weibull--von Mises model}

We use a Weibull distribution truncated on $(0,1)$ with the density as follows
\[
f_Q(q_t)
=
\frac{
\dfrac{r}{\lambda_{t|t-1}}
\left(\dfrac{q_t}{\lambda_{t|t-1}}\right)^{r-1}
\exp\!\left[
-\left(\dfrac{q_t}{\lambda_{t|t-1}}\right)^r
\right]
}{
1-\exp\!\left[-\lambda_{t|t-1}^{-r}\right]
},
\qquad 0<q_t<1,
\]
where $r>0$ is the shape parameter and
$
\lambda_{t|t-1}
=
\exp(\omega_Q+\eta_{t|t-1})>0
$
is the scale parameter.

Combining this range density with \eqref{eq:vm} gives the
centre--range truncated Weibull--von Mises (CR-TWvM) model.
Let $z_t=
\left(\frac{q_t}{\lambda_{t|t-1}}\right)^r,z_{1,t}=\lambda_{t|t-1}^{-r},$
the range score is
\begin{equation}
s_{Q,t}
=
r\left(
z_t-1+\frac{z_{1,t}}{\exp(z_{1,t})-1}
\right).
\label{WSC}
\end{equation}
The conditional mean is
\begin{equation}
    \mathbb E(Q_t\mid\mathcal F_{t-1})
=
\lambda_{t|t-1}
\frac{
\gamma\!\left(1+\frac{1}{r},z_{1,t}\right)
}{
1-\exp(-z_{1,t})
}, \label{TVMmean}
\end{equation}

where $\gamma(\cdot,\cdot)$ is the lower incomplete gamma function.

\subsubsection{Beta--von Mises model}
Next, we consider a reparameterized beta distribution with conditional mean
$m_{t|t-1}\in(0,1)$ and precision $\tau>0$
\citep{ferrari2004beta}. The density is
\[
f_Q(q_t)
=
\frac{
\Gamma(\tau)
}{
\Gamma(m_{t|t-1}\tau)
\Gamma\!\left((1-m_{t|t-1})\tau\right)
}
q_t^{m_{t|t-1}\tau-1}
(1-q_t)^{(1-m_{t|t-1})\tau-1},
\qquad 0<q_t<1,
\]
where the conditional mean is linked to the filter through
\[
m_{t|t-1}
=
\left\{
1+\exp(-\omega_Q-\eta_{t|t-1})
\right\}^{-1}.
\]

Combining this range density with \eqref{eq:vm} gives the
centre--range beta--von Mises (CR-BvM) model. Since
\[
\frac{\partial m_{t|t-1}}{\partial\eta_{t|t-1}}
=
m_{t|t-1}(1-m_{t|t-1}),
\]
the range score is
\begin{equation}
\begin{aligned}
s_{Q,t}
={}&
m_{t|t-1}(1-m_{t|t-1})\tau \times
\left[
\log\!\left(\frac{q_t}{1-q_t}\right)
-\psi^{*}(m_{t|t-1}\tau)
+\psi^{*}\!\left((1-m_{t|t-1})\tau\right)
\right],
\end{aligned}
\label{BSC}
\end{equation}
where $\psi^{*}(x)=\Gamma'(x)/\Gamma(x)$ is the digamma function.
The conditional mean is
\begin{equation}
    \mathbb E(Q_t\mid\mathcal F_{t-1})
=
m_{t|t-1}. \label{BVMMEAN}
\end{equation}

\subsubsection{Kumaraswamy--von Mises model}

Finally, we consider a Kumaraswamy distribution
\citep{kumaraswamy1980} with density
\[
f_Q(q_t)
=
a\,b_{t|t-1}\,
q_t^{a-1}
\left(1-q_t^a\right)^{b_{t|t-1}-1},
\qquad 0<q_t<1,
\]
where $a=\exp(\alpha)>0, b_{t|t-1}=\exp(\omega_Q+\eta_{t|t-1})>0
$ are shape parameters.

Combining this range density with \eqref{eq:vm} gives the
centre--range Kumaraswamy--von Mises (CR-KvM) model.
The range score is
\begin{equation}
s_{Q,t}
=
1+b_{t|t-1}\log\!\left(1-q_t^a\right).
\label{KSC}
\end{equation}
The conditional mean is
\begin{equation}
    \mathbb E(Q_t\mid\mathcal F_{t-1})
=
b_{t|t-1}
B\!\left(1+\frac{1}{a},b_{t|t-1}\right), \label{KVMMEAN}
\end{equation}

where $B(\cdot,\cdot)$ is the beta function.

\subsection{Score-driven dynamics for the centre--range}

Let $\mu_{t|t-1}$ describe changes in the location of the centre,
and let $\eta_{t|t-1}$ determine the time-varying parameter of the
range distribution. Following the score-driven framework of
\citet{harvey2024modelling,harvey2023regime}, define the filter vector
as
\[
\mathbf f_{t|t-1}
=
\begin{pmatrix}
\mu_{t|t-1}\\
\eta_{t|t-1}
\end{pmatrix}.
\]
The score is given by
\begin{equation}
\nabla_t
=
\frac{\partial\ell_t}{\partial\mathbf f_{t|t-1}}
=
\begin{pmatrix}
s_{C,t}\\
s_{Q,t}
\end{pmatrix},
\label{eq:score}
\end{equation}
where $s_{C,t}=
\upsilon_t(q_t)
\sin\!\left(c_t-\mu_C-\mu_{t|t-1}\right)$, and $s_{Q,t}$ is given by \eqref{WSC}, \eqref{BSC} or
\eqref{KSC}, depending on the range distribution.

Thus, the vector used to update
the filters is
$
\mathbf u_t
=
\begin{pmatrix}
s_{C,t}\\
s_{Q,t}
\end{pmatrix}
$.
The filters are updated according to
\begin{equation}
\mathbf f_{t+1|t}
=
\boldsymbol{\Phi}\mathbf f_{t|t-1}
+
\mathbf K\mathbf u_t,
\qquad
\mathbf f_{1|0}=\mathbf 0,
\label{eq:recursion}
\end{equation}
where
\[
\boldsymbol{\Phi}
=
\begin{pmatrix}
\phi_C&0\\
0&\phi_Q
\end{pmatrix},
\qquad
\mathbf K
=
\begin{pmatrix}
\kappa_C&0\\
0&\kappa_Q
\end{pmatrix}.
\]

\subsection{Maximum likelihood estimation and one-step-ahead prediction}

For each model, let the parameters be $
\boldsymbol{\psi}=\begin{pmatrix}
\boldsymbol{\psi}_C\\
\boldsymbol{\psi}_Q
\end{pmatrix},
$
where the centre parameter vector is
$
\boldsymbol{\psi}_C
=
(\kappa_C,\phi_C,\mu_C,\zeta,\delta)'
$ and the range parameter vector depends on the chosen distribution are
\begin{align*}
\text{CR-TWvM:}\qquad
\boldsymbol{\psi}_Q
&=
(\kappa_Q,\phi_Q,\omega_Q,\log r)',\\
\text{CR-BvM:}\qquad
\boldsymbol{\psi}_Q
&=
(\kappa_Q,\phi_Q,\omega_Q,\log\tau)',\\
\text{CR-KvM:}\qquad
\boldsymbol{\psi}_Q
&=
(\kappa_Q,\phi_Q,\omega_Q,\alpha)'.
\end{align*}
We estimate $r$, $\tau$ and $a$ on the logarithmic scale to ensure positivity. For observations $(c_t,q_t)$, $t=1,\ldots,T$, the conditional
log-likelihood is
\[
\ell_T(\boldsymbol{\psi})
=
\sum_{t=1}^{T}\ell_t(\boldsymbol{\psi})=\sum_{t=1}^{T}\log f_{C\mid Q,\mathcal F_{t-1}}(c_t\mid q_t)+\log f_Q(q_t\mid\mathcal F_{t-1})=\ell_T^C(\boldsymbol{\psi}_C)
+
\ell_T^Q(\boldsymbol{\psi}_Q),
\]
where
\[
\ell_T^C(\boldsymbol{\psi}_C)
=
\sum_{t=1}^{T}
\left[
\upsilon_t(q_t)
\cos\!\left(c_t-\mu_C-\mu_{t|t-1}\right)
-\log(2\pi)
-\log I_0\!\left(\upsilon_t(q_t)\right)
\right], 
\]
and for each linear distribution we have 
\begin{itemize}
 \item Truncated Weibull distribution
\[
\begin{aligned}
\ell_T^Q(\boldsymbol{\psi}_Q)
=
\sum_{t=1}^{T}
\Bigg[
&\log r-r\log\lambda_{t|t-1}
+(r-1)\log q_t-z_t
-\log\!\left(1-\exp(-z_{1,t})\right)
\Bigg],
\end{aligned}
\]
where $
\lambda_{t|t-1}=\exp(\omega_Q+\eta_{t|t-1}), z_t=
\left(\frac{q_t}{\lambda_{t|t-1}}\right)^r,
z_{1,t}=\lambda_{t|t-1}^{-r}.$

\item Beta distribution
\[
\begin{aligned}
\ell_T^Q(\boldsymbol{\psi}_Q)
=
\sum_{t=1}^{T}
\Big[
&\log\Gamma(\tau)
-\log\Gamma(m_{t|t-1}\tau)-\log\Gamma\!\left((1-m_{t|t-1})\tau\right)+(m_{t|t-1}\tau-1)\log q_t\\
&+\left((1-m_{t|t-1})\tau-1\right)\log(1-q_t)
\Big],
\end{aligned}
\]
where
$m_{t|t-1}=\left\{1+\exp(-\omega_Q-\eta_{t|t-1})\right\}^{-1}.
$

\item Kumaraswamy distribution
\[
\begin{aligned}
\ell_T^Q(\boldsymbol{\psi}_Q)
=
\sum_{t=1}^{T}
\Big[\log a+\log b_{t|t-1}
+(a-1)\log q_t+(b_{t|t-1}-1)\log(1-q_t^a)
\Big],
\end{aligned}
\]
where $a=\exp(\alpha),~
 b_{t|t-1}
=
\exp(\omega_Q+\eta_{t|t-1}).
$
\end{itemize}
The maximum likelihood estimator is
\[
\widehat{\boldsymbol{\psi}}
=
\mathop{\operatorname{arg\,max}}_{\boldsymbol{\psi}}
\ell_T(\boldsymbol{\psi}).
\]
The log-likelihood is maximised numerically.
After observing $(c_t,q_t)$, the filters are updated to obtain
$\mu_{t+1|t}$ and $\eta_{t+1|t}$. The one-step-ahead location
forecast of the centre is
\[
\widehat C_{t+1|t}
=
\operatorname{mod}_{2\pi}
\left(\widehat\mu_C+\widehat\mu_{t+1|t}+\pi\right)-\pi.
\]
The range forecast is obtained as
\[
\widehat Q_{t+1|t}
=
\widehat{\mathbb E}(Q_{t+1}\mid\mathcal F_t),
\]
calculated using the conditional mean in \ref{TVMmean}, \ref{BVMMEAN} and \ref{KVMMEAN} for the chosen
range distribution.

\section{Simulation Study}
\label{sec:simulation} 
In this section, a simulation study is performed to assess the performance of the maximum
likelihood estimators of CR-WvM, CR-BvM and CR-KvM. For each model, data are generated under four scenarios
that differ in the location of the centre, the variability of the range, the persistence of
the score dynamics and the circular-linear dependence. Performance is assessed using the absolute bias (ABIAS), standard deviation (SD) and mean squared error (MSE) of the
estimators. For each model and scenario, 1000 replications of samples of sizes $T \in \{100, 500, 1000\}$ are generated from the three models mentioned above.

For model $k \in \{\text{CR-TWvM}, \text{CR-BvM}, \text{CR-KvM}\}$ and a given scenario, let
$\boldsymbol{\psi}^{(k)} = \bigl(\psi^{(k)}_1, \ldots, \psi^{(k)}_9\bigr)'
$ denote the vector of true parameter values, containing $\boldsymbol{\psi}_c$ and $\boldsymbol{\psi}_Q$.

\begin{figure}[p]
    \centering
    \includegraphics[angle=90, height=\dimexpr\textheight-5\baselineskip\relax]{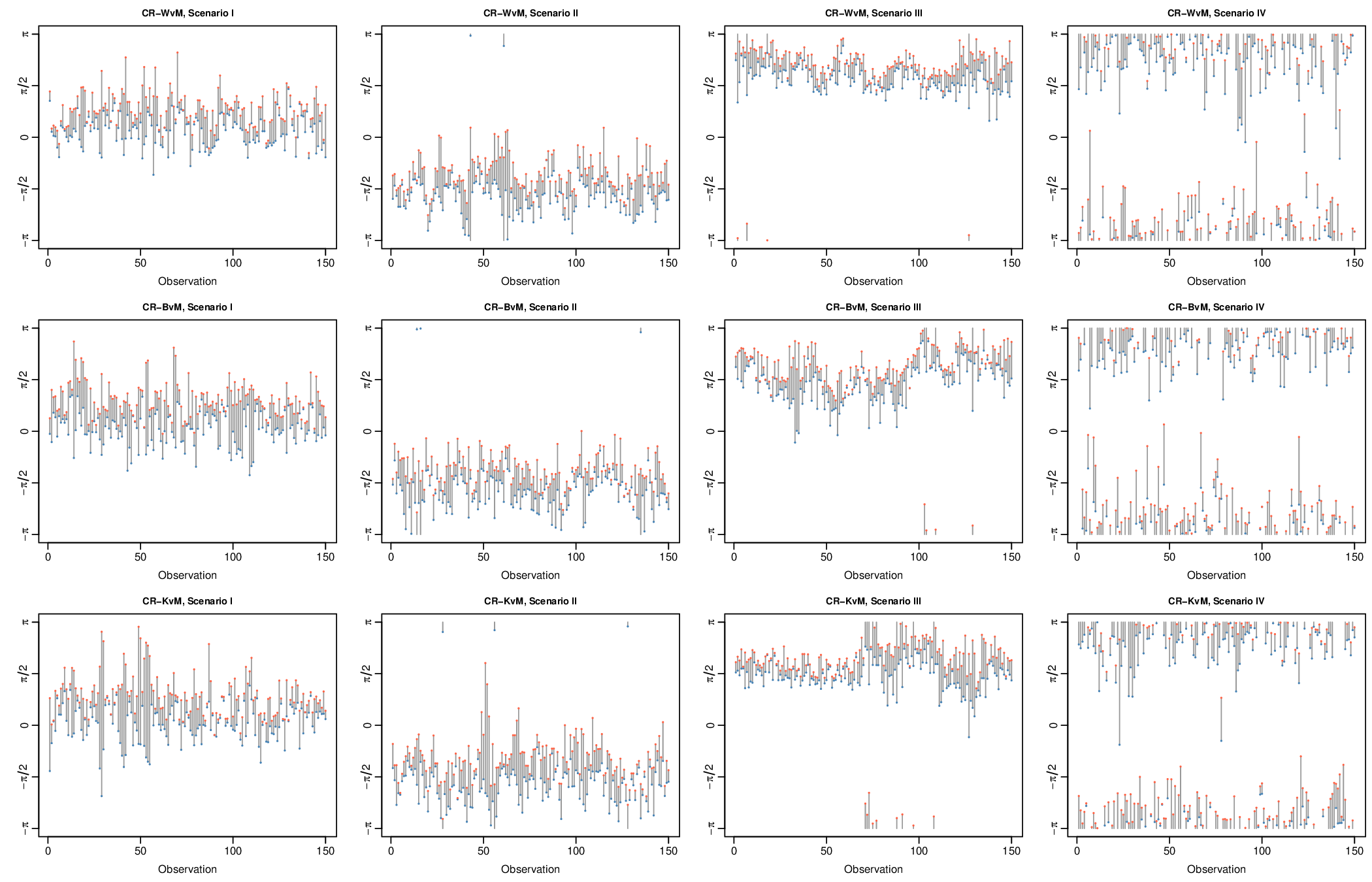}
    \caption{Simulated interval-valued series from the CR-WvM (top row), CR-BvM (middle row), and
    CR-KvM (bottom row) models under Scenarios I--IV (left to right), with $T = 150$.}
    \label{fig:sim_series}
\end{figure}

For each distribution, we consider four simulation scenarios covering narrow and wide arcs, low, moderate and high dependence between the centre and range, and different patterns of temporal dynamics. The scenarios also include intervals that cross the circular cut (see Figure \ref{fig:sim_series}).
In all three models, $\boldsymbol{\psi}_C$ is the same. The full parameter vectors are as follows.
\begin{itemize}
  \item \textbf{CR-TWvM:}
        $\boldsymbol{\psi} = (\omega_C, \phi_C, \kappa_C, \omega_Q, \phi_Q, \kappa_Q, \log r, \zeta, \delta)'$.
  \begin{itemize}
    \item Scenario I:   $\boldsymbol{\psi} = (0.5, 0.30, 0.40, -2.204, 0.3, 0.10, 0.701, 2.3, 0.5)'$
    \item Scenario II:  $\boldsymbol{\psi} = (-1.5, 0.60, 0.35, -2.023, 0.6, 0.16, 0.442, 2.0, 2.0)'$
    \item Scenario III: $\boldsymbol{\psi} = (2.0, 0.95, 0.50, -2.403, 0.9, 0.05, 0.698, 2.8, 1.0)'$
    \item Scenario IV:  $\boldsymbol{\psi} = (3.0, 0.60, 0.25, -1.895, 0.6, 0.16, 0.450, 1.6, 4.0)'$
  \end{itemize}

  \item \textbf{CR-BvM:}
        $\boldsymbol{\psi} = (\omega_C, \phi_C, \kappa_C, \omega_Q, \phi_Q, \kappa_Q, \log\tau, \zeta, \delta)'$.
  \begin{itemize}
    \item Scenario I:   $\boldsymbol{\psi} = (0.5, 0.30, 0.40, -2.223, 0.3, 0.11, 3.505, 2.3, 0.5)'$
    \item Scenario II:  $\boldsymbol{\psi} = (-1.5, 0.60, 0.35, -2.003, 0.6, 0.17, 2.785, 2.0, 2.0)'$
    \item Scenario III: $\boldsymbol{\psi} = (2.0, 0.95, 0.50, -2.440, 0.9, 0.05, 3.723, 2.8, 1.0)'$
    \item Scenario IV:  $\boldsymbol{\psi} = (3.0, 0.60, 0.25, -1.857, 0.6, 0.17, 2.645, 1.6, 4.0)'$
  \end{itemize}

  \item \textbf{CR-KvM:}
        $\boldsymbol{\psi} = (\omega_C, \phi_C, \kappa_C, \omega_Q, \phi_Q, \kappa_Q, \alpha, \zeta, \delta)'$.
  \begin{itemize}
    \item Scenario I:   $\boldsymbol{\psi} = (0.5, 0.30, 0.40, 4.4, 0.3, 0.20, \log 2, 2.3, 0.5)'$
    \item Scenario II:  $\boldsymbol{\psi} = (-1.5, 0.60, 0.35, 3.0, 0.6, 0.25, \log 1.5, 2.0, 2.0)'$
    \item Scenario III: $\boldsymbol{\psi} = (2.0, 0.95, 0.50, 4.8, 0.9, 0.10, \log 2, 2.8, 1.0)'$
    \item Scenario IV:  $\boldsymbol{\psi} = (3.0, 0.60, 0.25, 2.8, 0.6, 0.25, \log 1.5, 1.6, 4.0)'$
  \end{itemize}
\end{itemize}

The results of the parameter estimates, ABIAS, SD and MSE are provided in Tables \ref{tab:wvm_sim}-\ref{tab:kvm_sim} for scenarios I to IV.

Additionally, we also consider various settings of the circular-linear dependence parameter, $\delta$. Four
settings are provided ($\delta = 0, \delta = 2, \delta = -2, \delta = 5$).  Tables~\ref{tab:wvm_delta}--\ref{tab:kvm_delta} report the parameter
estimates along with ABIAS, SD and MSE for the CR-TWvM, CR-BvM and CR-KvM models. The maximum likelihood estimators perform well across all simulation scenarios. The estimation accuracy improves as the sample size increases, with the ABIAS and MSE decreasing. In conclusion, these findings suggest that the proposed estimation procedure performs well and provides accurate estimates.

\begin{table}[H]
\centering
\caption{Parameter estimates, ABIAS, SD and MSE for the CR-TWvM model.}
\label{tab:wvm_sim}
\scriptsize
\setlength{\tabcolsep}{3pt}
\renewcommand{\arraystretch}{1.1}
\begin{subtable}[t]{0.49\textwidth}
\centering
\begin{adjustbox}{max width=\linewidth}
\begin{tabular}{|c c S[table-format=-1.4] S[table-format=-1.4]
                    S[table-format=<1.4] S[table-format=1.4] S[table-format=<2.4]|}
\hline
$T$ & Parameter & {True} & {Estimate} & {ABIAS} & {SD} & {MSE} \\
\hline
\multicolumn{7}{|c|}{I} \\
\hline
\multirow{9}{*}{100} & $\omega_C$ & 0.5000 & 0.5018 & 0.0018 & 0.0484 & 0.0023 \\
 & $\phi_C$ & 0.3000 & 0.2086 & 0.0914 & 0.3055 & 0.1015 \\
 & $\kappa_C$ & 0.4000 & 0.3824 & 0.0176 & 0.1104 & 0.0125 \\
 & $\omega_Q$ & -2.0000 & -2.0125 & 0.0125 & 0.1250 & 0.0157 \\
 & $\phi_Q$ & 0.3000 & 0.2034 & 0.0966 & 0.3968 & 0.1665 \\
 & $\kappa_Q$ & 0.1700 & 0.1642 & 0.0058 & 0.0747 & 0.0056 \\
 & $\log\tau$ & 2.1270 & 2.1823 & 0.0553 & 0.1539 & 0.0267 \\
 & $\zeta$ & 2.3000 & 2.3328 & 0.0328 & 0.2132 & 0.0464 \\
 & $\delta$ & 0.5000 & 0.3956 & 0.1044 & 1.3871 & 1.9310 \\
\cline{1-7}
\multirow{9}{*}{500} & $\omega_C$ & 0.5000 & 0.5005 & 0.0005 & 0.0234 & 0.0005 \\
 & $\phi_C$ & 0.3000 & 0.2815 & 0.0185 & 0.1208 & 0.0149 \\
 & $\kappa_C$ & 0.4000 & 0.3935 & 0.0065 & 0.0481 & 0.0023 \\
 & $\omega_Q$ & -2.0000 & -2.0002 & 0.0002 & 0.0570 & 0.0032 \\
 & $\phi_Q$ & 0.3000 & 0.2687 & 0.0313 & 0.1935 & 0.0384 \\
 & $\kappa_Q$ & 0.1700 & 0.1678 & 0.0022 & 0.0346 & 0.0012 \\
 & $\log\tau$ & 2.1270 & 2.1350 & 0.0080 & 0.0670 & 0.0045 \\
 & $\zeta$ & 2.3000 & 2.3009 & 0.0009 & 0.0924 & 0.0085 \\
 & $\delta$ & 0.5000 & 0.4550 & 0.0450 & 0.5882 & 0.3473 \\
\cline{1-7}
\multirow{9}{*}{1000} & $\omega_C$ & 0.5000 & 0.5002 & 0.0002 & 0.0164 & 0.0003 \\
 & $\phi_C$ & 0.3000 & 0.2920 & 0.0080 & 0.0883 & 0.0079 \\
 & $\kappa_C$ & 0.4000 & 0.3974 & 0.0026 & 0.0338 & 0.0011 \\
 & $\omega_Q$ & -2.0000 & -2.0027 & 0.0027 & 0.0389 & 0.0015 \\
 & $\phi_Q$ & 0.3000 & 0.2890 & 0.0110 & 0.1186 & 0.0142 \\
 & $\kappa_Q$ & 0.1700 & 0.1688 & 0.0012 & 0.0238 & 0.0006 \\
 & $\log\tau$ & 2.1270 & 2.1319 & 0.0049 & 0.0459 & 0.0021 \\
 & $\zeta$ & 2.3000 & 2.3057 & 0.0057 & 0.0647 & 0.0042 \\
 & $\delta$ & 0.5000 & 0.5089 & 0.0089 & 0.3941 & 0.1550 \\
\hline
\multicolumn{7}{|c|}{II} \\
\hline
\multirow{9}{*}{100} & $\omega_C$ & -1.5000 & -1.4972 & 0.0028 & 0.0779 & 0.0061 \\
 & $\phi_C$ & 0.6000 & 0.4771 & 0.1229 & 0.3229 & 0.1191 \\
 & $\kappa_C$ & 0.3500 & 0.3286 & 0.0214 & 0.1185 & 0.0145 \\
 & $\omega_Q$ & -2.0230 & -2.0328 & 0.0098 & 0.1329 & 0.0177 \\
 & $\phi_Q$ & 0.6000 & 0.5211 & 0.0789 & 0.2453 & 0.0662 \\
 & $\kappa_Q$ & 0.1600 & 0.1453 & 0.0147 & 0.0507 & 0.0028 \\
 & $\log r$ & 0.4420 & 0.4699 & 0.0279 & 0.0811 & 0.0073 \\
 & $\zeta$ & 2.0000 & 2.0183 & 0.0183 & 0.2374 & 0.0565 \\
 & $\delta$ & 2.0000 & 1.7824 & 0.2176 & 1.6982 & 2.9239 \\
\cline{1-7}
\multirow{9}{*}{500} & $\omega_C$ & -1.5000 & -1.5012 & 0.0012 & 0.0347 & 0.0012 \\
 & $\phi_C$ & 0.6000 & 0.5877 & 0.0123 & 0.0941 & 0.0090 \\
 & $\kappa_C$ & 0.3500 & 0.3458 & 0.0042 & 0.0459 & 0.0021 \\
 & $\omega_Q$ & -2.0230 & -2.0206 & 0.0024 & 0.0617 & 0.0038 \\
 & $\phi_Q$ & 0.6000 & 0.5881 & 0.0119 & 0.0811 & 0.0067 \\
 & $\kappa_Q$ & 0.1600 & 0.1578 & 0.0022 & 0.0220 & 0.0005 \\
 & $\log r$ & 0.4420 & 0.4467 & 0.0047 & 0.0335 & 0.0011 \\
 & $\zeta$ & 2.0000 & 2.0115 & 0.0115 & 0.0944 & 0.0090 \\
 & $\delta$ & 2.0000 & 2.0406 & 0.0406 & 0.6189 & 0.3837 \\
\cline{1-7}
\multirow{9}{*}{1000} & $\omega_C$ & -1.5000 & -1.4992 & 0.0008 & 0.0244 & 0.0006 \\
 & $\phi_C$ & 0.6000 & 0.5927 & 0.0073 & 0.0659 & 0.0044 \\
 & $\kappa_C$ & 0.3500 & 0.3469 & 0.0031 & 0.0314 & 0.0010 \\
 & $\omega_Q$ & -2.0230 & -2.0227 & 0.0003 & 0.0405 & 0.0016 \\
 & $\phi_Q$ & 0.6000 & 0.5946 & 0.0054 & 0.0566 & 0.0032 \\
 & $\kappa_Q$ & 0.1600 & 0.1608 & 0.0008 & 0.0147 & 0.0002 \\
 & $\log r$ & 0.4420 & 0.4418 & 0.0002 & 0.0240 & 0.0006 \\
 & $\zeta$ & 2.0000 & 2.0074 & 0.0074 & 0.0653 & 0.0043 \\
 & $\delta$ & 2.0000 & 2.0175 & 0.0175 & 0.4126 & 0.1701 \\
\hline
\end{tabular}
\end{adjustbox}
\end{subtable}%
\begin{subtable}[t]{0.49\textwidth}
\centering
\begin{adjustbox}{max width=\linewidth}
\begin{tabular}{|c c S[table-format=-1.4] S[table-format=-1.4]
                    S[table-format=<1.4] S[table-format=1.4] S[table-format=<2.4]|}
\hline
$T$ & Parameter & {True} & {Estimate} & {ABIAS} & {SD} & {MSE} \\
\hline
\multicolumn{7}{|c|}{III} \\
\hline
\multirow{9}{*}{100} & $\omega_C$ & -1.5000 & -1.4994 & 0.0006 & 0.0780 & 0.0061 \\
 & $\phi_C$ & 0.6000 & 0.4809 & 0.1191 & 0.3077 & 0.1087 \\
 & $\kappa_C$ & 0.3500 & 0.3317 & 0.0183 & 0.1091 & 0.0122 \\
 & $\omega_Q$ & -2.0030 & -2.0056 & 0.0026 & 0.1477 & 0.0218 \\
 & $\phi_Q$ & 0.6000 & 0.5223 & 0.0777 & 0.2262 & 0.0571 \\
 & $\kappa_Q$ & 0.1700 & 0.1635 & 0.0065 & 0.0496 & 0.0025 \\
 & $\log\tau$ & 2.7850 & 2.8322 & 0.0472 & 0.1550 & 0.0262 \\
 & $\zeta$ & 2.0000 & 2.0305 & 0.0305 & 0.2506 & 0.0636 \\
 & $\delta$ & 2.0000 & 1.8074 & 0.1926 & 1.6607 & 2.7897 \\
\cline{1-7}
\multirow{9}{*}{500} & $\omega_C$ & -1.5000 & -1.5026 & 0.0026 & 0.0353 & 0.0013 \\
 & $\phi_C$ & 0.6000 & 0.5797 & 0.0203 & 0.1054 & 0.0115 \\
 & $\kappa_C$ & 0.3500 & 0.3458 & 0.0042 & 0.0497 & 0.0025 \\
 & $\omega_Q$ & -2.0030 & -2.0020 & 0.0010 & 0.0569 & 0.0032 \\
 & $\phi_Q$ & 0.6000 & 0.5788 & 0.0212 & 0.0897 & 0.0085 \\
 & $\kappa_Q$ & 0.1700 & 0.1683 & 0.0017 & 0.0215 & 0.0005 \\
 & $\log\tau$ & 2.7850 & 2.7979 & 0.0129 & 0.0659 & 0.0045 \\
 & $\zeta$ & 2.0000 & 2.0064 & 0.0064 & 0.1048 & 0.0110 \\
 & $\delta$ & 2.0000 & 1.9542 & 0.0458 & 0.6913 & 0.4790 \\
\cline{1-7}
\multirow{9}{*}{1000} & $\omega_C$ & -1.5000 & -1.5001 & 0.0001 & 0.0254 & 0.0006 \\
 & $\phi_C$ & 0.6000 & 0.5919 & 0.0081 & 0.0674 & 0.0046 \\
 & $\kappa_C$ & 0.3500 & 0.3500 & <0.0001 & 0.0359 & 0.0013 \\
 & $\omega_Q$ & -2.0030 & -2.0052 & 0.0022 & 0.0432 & 0.0019 \\
 & $\phi_Q$ & 0.6000 & 0.5868 & 0.0132 & 0.0603 & 0.0038 \\
 & $\kappa_Q$ & 0.1700 & 0.1691 & 0.0009 & 0.0155 & 0.0002 \\
 & $\log\tau$ & 2.7850 & 2.7909 & 0.0059 & 0.0475 & 0.0023 \\
 & $\zeta$ & 2.0000 & 2.0093 & 0.0093 & 0.0736 & 0.0055 \\
 & $\delta$ & 2.0000 & 2.0234 & 0.0234 & 0.4735 & 0.2243 \\
\hline
\multicolumn{7}{|c|}{IV} \\
\hline
\multirow{9}{*}{100} & $\omega_C$ & 3.0000 & 3.0027 & 0.0027 & 0.1114 & 0.0124 \\
 & $\phi_C$ & 0.6000 & 0.4442 & 0.1558 & 0.4051 & 0.1880 \\
 & $\kappa_C$ & 0.2500 & 0.2305 & 0.0195 & 0.1391 & 0.0197 \\
 & $\omega_Q$ & -1.8950 & -1.8776 & 0.0174 & 0.1412 & 0.0202 \\
 & $\phi_Q$ & 0.6000 & 0.5349 & 0.0651 & 0.2348 & 0.0592 \\
 & $\kappa_Q$ & 0.1600 & 0.1505 & 0.0095 & 0.0536 & 0.0030 \\
 & $\log r$ & 0.4500 & 0.4736 & 0.0236 & 0.0801 & 0.0070 \\
 & $\zeta$ & 1.6000 & 1.6654 & 0.0654 & 0.2393 & 0.0614 \\
 & $\delta$ & 4.0000 & 4.1341 & 0.1341 & 1.5815 & 2.5130 \\
\cline{1-7}
\multirow{9}{*}{500} & $\omega_C$ & 3.0000 & 3.0020 & 0.0020 & 0.0388 & 0.0015 \\
 & $\phi_C$ & 0.6000 & 0.5647 & 0.0353 & 0.1448 & 0.0222 \\
 & $\kappa_C$ & 0.2500 & 0.2484 & 0.0016 & 0.0547 & 0.0030 \\
 & $\omega_Q$ & -1.8950 & -1.8994 & 0.0044 & 0.0562 & 0.0032 \\
 & $\phi_Q$ & 0.6000 & 0.5889 & 0.0111 & 0.0782 & 0.0062 \\
 & $\kappa_Q$ & 0.1600 & 0.1598 & 0.0002 & 0.0236 & 0.0006 \\
 & $\log r$ & 0.4500 & 0.4515 & 0.0015 & 0.0370 & 0.0014 \\
 & $\zeta$ & 1.6000 & 1.6102 & 0.0102 & 0.1070 & 0.0115 \\
 & $\delta$ & 4.0000 & 4.0177 & 0.0177 & 0.6517 & 0.4239 \\
\cline{1-7}
\multirow{9}{*}{1000} & $\omega_C$ & 3.0000 & 3.0018 & 0.0018 & 0.0309 & 0.0010 \\
 & $\phi_C$ & 0.6000 & 0.5786 & 0.0214 & 0.1047 & 0.0114 \\
 & $\kappa_C$ & 0.2500 & 0.2516 & 0.0016 & 0.0374 & 0.0014 \\
 & $\omega_Q$ & -1.8950 & -1.8949 & 0.0001 & 0.0398 & 0.0016 \\
 & $\phi_Q$ & 0.6000 & 0.5980 & 0.0020 & 0.0536 & 0.0029 \\
 & $\kappa_Q$ & 0.1600 & 0.1588 & 0.0012 & 0.0155 & 0.0002 \\
 & $\log r$ & 0.4500 & 0.4541 & 0.0041 & 0.0252 & 0.0007 \\
 & $\zeta$ & 1.6000 & 1.6071 & 0.0071 & 0.0731 & 0.0054 \\
 & $\delta$ & 4.0000 & 4.0369 & 0.0369 & 0.4643 & 0.2164 \\
\hline
\end{tabular}
\end{adjustbox}
\end{subtable}
\end{table}

\begin{table}[H]
\centering
\caption{Parameter estimates, ABIAS, SD and MSE for the CR-BvM model.}
\label{tab:bvm_sim}
\scriptsize
\setlength{\tabcolsep}{3pt}
\renewcommand{\arraystretch}{1.1}
\begin{subtable}[t]{0.49\textwidth}
\centering
\begin{adjustbox}{max width=\linewidth}
\begin{tabular}{|c c S[table-format=-1.4] S[table-format=-1.4]
                    S[table-format=<1.4] S[table-format=1.4] S[table-format=<2.4]|}
\hline
$T$ & Parameter & {True} & {Estimate} & {ABIAS} & {SD} & {MSE} \\
\hline
\multicolumn{7}{|c|}{I} \\
\hline
\multirow{9}{*}{100} & $\omega_C$ & 0.5000 & 0.5007 & 0.0007 & 0.0532 & 0.0028 \\
 & $\phi_C$ & 0.3000 & 0.2089 & 0.0911 & 0.2951 & 0.0952 \\
 & $\kappa_C$ & 0.4000 & 0.3845 & 0.0155 & 0.1079 & 0.0119 \\
 & $\omega_Q$ & -2.2230 & -2.2235 & 0.0005 & 0.0829 & 0.0069 \\
 & $\phi_Q$ & 0.3000 & 0.2407 & 0.0593 & 0.2753 & 0.0791 \\
 & $\kappa_Q$ & 0.1100 & 0.1030 & 0.0070 & 0.0340 & 0.0012 \\
 & $\log\tau$ & 3.5050 & 3.5456 & 0.0406 & 0.1481 & 0.0235 \\
 & $\zeta$ & 2.3000 & 2.3552 & 0.0552 & 0.3171 & 0.1033 \\
 & $\delta$ & 0.5000 & 0.4691 & 0.0309 & 1.3672 & 1.9763\\
\cline{1-7}
\multirow{9}{*}{500} & $\omega_C$ & 0.5000 & 0.5037 & 0.0037 & 0.0239 & 0.0006 \\
 & $\phi_C$ & 0.3000 & 0.2922 & 0.0078 & 0.1173 & 0.0138 \\
 & $\kappa_C$ & 0.4000 & 0.3952 & 0.0048 & 0.0476 & 0.0023 \\
 & $\omega_Q$ & -2.2230 & -2.2251 & 0.0021 & 0.0395 & 0.0016 \\
 & $\phi_Q$ & 0.3000 & 0.2944 & 0.0056 & 0.1196 & 0.0143 \\
 & $\kappa_Q$ & 0.1100 & 0.1086 & 0.0014 & 0.0153 & 0.0002 \\
 & $\log\tau$ & 3.5050 & 3.5144 & 0.0094 & 0.0618 & 0.0039 \\
 & $\zeta$ & 2.3000 & 2.3170 & 0.0170 & 0.1359 & 0.0187 \\
 & $\delta$ & 0.5000 & 0.5676 & 0.0676 & 1.1893 & 1.4156 \\
\cline{1-7}
\multirow{9}{*}{1000} & $\omega_C$ & 0.5000 & 0.4995 & 0.0005 & 0.0157 & 0.0002 \\
 & $\phi_C$ & 0.3000 & 0.2945 & 0.0055 & 0.0843 & 0.0071 \\
 & $\kappa_C$ & 0.4000 & 0.3969 & 0.0031 & 0.0337 & 0.0011 \\
 & $\omega_Q$ & -2.2230 & -2.2248 & 0.0018 & 0.0279 & 0.0008 \\
 & $\phi_Q$ & 0.3000 & 0.2933 & 0.0067 & 0.0861 & 0.0074 \\
 & $\kappa_Q$ & 0.1100 & 0.1082 & 0.0018 & 0.0103 & 0.0001 \\
 & $\log\tau$ & 3.5050 & 3.5122 & 0.0072 & 0.0449 & 0.0021 \\
 & $\zeta$ & 2.3000 & 2.3112 & 0.0112 & 0.0845 & 0.0072 \\
 & $\delta$ & 0.5000 & 0.5553 & 0.0553 & 0.7643 & 0.5858 \\
\hline
\multicolumn{7}{|c|}{II} \\
\hline
\multirow{9}{*}{100} & $\omega_C$ & -1.5000 & -1.5002 & 0.0002 & 0.0784 & 0.0061 \\
 & $\phi_C$ & 0.6000 & 0.4832 & 0.1168 & 0.3087 & 0.1087 \\
 & $\kappa_C$ & 0.3500 & 0.3340 & 0.0160 & 0.1100 & 0.0123 \\
 & $\omega_Q$ & -2.0030 & -1.9980 & 0.0050 & 0.1470 & 0.0216 \\
 & $\phi_Q$ & 0.6000 & 0.5205 & 0.0795 & 0.2192 & 0.0542 \\
 & $\kappa_Q$ & 0.1700 & 0.1639 & 0.0061 & 0.0500 & 0.0025 \\
 & $\log\tau$ & 2.7850 & 2.8297 & 0.0447 & 0.1573 & 0.0267 \\
 & $\zeta$ & 2.0000 & 2.0327 & 0.0327 & 0.2477 & 0.0623 \\
 & $\delta$ & 2.0000 & 1.8381 & 0.1619 & 1.6433 & 2.7199 \\
\cline{1-7}
\multirow{9}{*}{500} & $\omega_C$ & -1.5000 & -1.5022 & 0.0022 & 0.0358 & 0.0013 \\
 & $\phi_C$ & 0.6000 & 0.5764 & 0.0236 & 0.1067 & 0.0119 \\
 & $\kappa_C$ & 0.3500 & 0.3465 & 0.0035 & 0.0490 & 0.0024 \\
 & $\omega_Q$ & -2.0030 & -2.0026 & 0.0004 & 0.0584 & 0.0034 \\
 & $\phi_Q$ & 0.6000 & 0.5769 & 0.0231 & 0.0892 & 0.0085 \\
 & $\kappa_Q$ & 0.1700 & 0.1678 & 0.0022 & 0.0218 & 0.0005 \\
 & $\log\tau$ & 2.7850 & 2.8003 & 0.0153 & 0.0646 & 0.0044 \\
 & $\zeta$ & 2.0000 & 2.0017 & 0.0017 & 0.1061 & 0.0112 \\
 & $\delta$ & 2.0000 & 1.9140 & 0.0860 & 0.7064 & 0.5052 \\
\cline{1-7}
\multirow{9}{*}{1000} & $\omega_C$ & -1.5000 & -1.5001 & 0.0001 & 0.0263 & 0.0007 \\
 & $\phi_C$ & 0.6000 & 0.5901 & 0.0099 & 0.0687 & 0.0048 \\
 & $\kappa_C$ & 0.3500 & 0.3495 & 0.0005 & 0.0362 & 0.0013 \\
 & $\omega_Q$ & -2.0030 & -2.0066 & 0.0036 & 0.0419 & 0.0018 \\
 & $\phi_Q$ & 0.6000 & 0.5861 & 0.0139 & 0.0618 & 0.0040 \\
 & $\kappa_Q$ & 0.1700 & 0.1697 & 0.0003 & 0.0157 & 0.0002 \\
 & $\log\tau$ & 2.7850 & 2.7923 & 0.0073 & 0.0466 & 0.0022 \\
 & $\zeta$ & 2.0000 & 2.0121 & 0.0121 & 0.0734 & 0.0055 \\
 & $\delta$ & 2.0000 & 2.0378 & 0.0378 & 0.4770 & 0.2284 \\
\hline
\end{tabular}
\end{adjustbox}
\end{subtable}%
\begin{subtable}[t]{0.49\textwidth}
\centering
\begin{adjustbox}{max width=\linewidth}
\begin{tabular}{|c c S[table-format=-1.4] S[table-format=-1.4]
                    S[table-format=<1.4] S[table-format=1.4] S[table-format=<2.4]|}
\hline
$T$ & Parameter & {True} & {Estimate} & {ABIAS} & {SD} & {MSE} \\
\hline
\multicolumn{7}{|c|}{III} \\
\hline
\multirow{9}{*}{100} & $\omega_C$ & 2.0000 & 1.9831 & 0.0169 & 0.3683 & 0.1356 \\
 & $\phi_C$ & 0.9500 & 0.9078 & 0.0422 & 0.0860 & 0.0092 \\
 & $\kappa_C$ & 0.5000 & 0.4899 & 0.0101 & 0.0981 & 0.0097 \\
 & $\omega_Q$ & -2.4400 & -2.4430 & 0.0030 & 0.1891 & 0.0357 \\
 & $\phi_Q$ & 0.9000 & 0.7722 & 0.1278 & 0.2421 & 0.0748 \\
 & $\kappa_Q$ & 0.0500 & 0.0454 & 0.0046 & 0.0268 & 0.0007 \\
 & $\log\tau$ & 3.7230 & 3.7715 & 0.0485 & 0.1412 & 0.0222 \\
 & $\zeta$ & 2.8000 & 2.8294 & 0.0294 & 0.3103 & 0.0969 \\
 & $\delta$ & 1.0000 & 0.7389 & 0.2611 & 1.4562 & 1.9876 \\
\cline{1-7}
\multirow{9}{*}{500} & $\omega_C$ & 2.0000 & 1.9867 & 0.0133 & 0.2516 & 0.0633 \\
 & $\phi_C$ & 0.9500 & 0.9460 & 0.0040 & 0.0224 & 0.0005 \\
 & $\kappa_C$ & 0.5000 & 0.5000 & <0.0001 & 0.0421 & 0.0018 \\
 & $\omega_Q$ & -2.4400 & -2.4374 & 0.0026 & 0.0646 & 0.0042 \\
 & $\phi_Q$ & 0.9000 & 0.8840 & 0.0160 & 0.0479 & 0.0025 \\
 & $\kappa_Q$ & 0.0500 & 0.0499 & 0.0001 & 0.0102 & 0.0001 \\
 & $\log\tau$ & 3.7230 & 3.7296 & 0.0066 & 0.0648 & 0.0042 \\
 & $\zeta$ & 2.8000 & 2.8005 & 0.0005 & 0.1236 & 0.0152 \\
 & $\delta$ & 1.0000 & 0.9226 & 0.0774 & 1.3273 & 1.7633 \\
\cline{1-7}
\multirow{9}{*}{1000} & $\omega_C$ & 2.0000 & 1.9909 & 0.0091 & 0.1128 & 0.0128 \\
 & $\phi_C$ & 0.9500 & 0.9459 & 0.0041 & 0.0124 & 0.0002 \\
 & $\kappa_C$ & 0.5000 & 0.4999 & 0.0001 & 0.0301 & 0.0009 \\
 & $\omega_Q$ & -2.4400 & -2.4426 & 0.0026 & 0.0470 & 0.0022 \\
 & $\phi_Q$ & 0.9000 & 0.8898 & 0.0102 & 0.0310 & 0.0011 \\
 & $\kappa_Q$ & 0.0500 & 0.0500 & <0.0001 & 0.0070 & <0.0001 \\
 & $\log\tau$ & 3.7230 & 3.7249 & 0.0019 & 0.0472 & 0.0022 \\
 & $\zeta$ & 2.8000 & 2.8037 & 0.0037 & 0.0872 & 0.0076 \\
 & $\delta$ & 1.0000 & 1.0159 & 0.0159 & 0.9773 & 0.9530 \\
\hline
\multicolumn{7}{|c|}{IV} \\
\hline
\multirow{9}{*}{100} & $\omega_C$ & 3.0000 & 2.9962 & 0.0038 & 0.0977 & 0.0095 \\
 & $\phi_C$ & 0.6000 & 0.4301 & 0.1699 & 0.3914 & 0.1817 \\
 & $\kappa_C$ & 0.2500 & 0.2407 & 0.0093 & 0.1296 & 0.0168 \\
 & $\omega_Q$ & -1.8570 & -1.8693 & 0.0123 & 0.1331 & 0.0178 \\
 & $\phi_Q$ & 0.6000 & 0.5126 & 0.0874 & 0.2686 & 0.0796 \\
 & $\kappa_Q$ & 0.1700 & 0.1547 & 0.0153 & 0.0498 & 0.0027 \\
 & $\log\tau$ & 2.6450 & 2.6955 & 0.0505 & 0.1450 & 0.0235 \\
 & $\zeta$ & 1.6000 & 1.6947 & 0.0947 & 0.2412 & 0.0670 \\
 & $\delta$ & 4.0000 & 4.3337 & 0.3337 & 1.5825 & 2.6093 \\
\cline{1-7}
\multirow{9}{*}{500} & $\omega_C$ & 3.0000 & 2.9956 & 0.0044 & 0.0427 & 0.0018 \\
 & $\phi_C$ & 0.6000 & 0.5729 & 0.0271 & 0.1389 & 0.0200 \\
 & $\kappa_C$ & 0.2500 & 0.2492 & 0.0008 & 0.0516 & 0.0027 \\
 & $\omega_Q$ & -1.8570 & -1.8566 & 0.0004 & 0.0610 & 0.0037 \\
 & $\phi_Q$ & 0.6000 & 0.5911 & 0.0089 & 0.0902 & 0.0082 \\
 & $\kappa_Q$ & 0.1700 & 0.1695 & 0.0005 & 0.0221 & 0.0005 \\
 & $\log\tau$ & 2.6450 & 2.6526 & 0.0076 & 0.0643 & 0.0042 \\
 & $\zeta$ & 1.6000 & 1.6199 & 0.0199 & 0.1089 & 0.0122 \\
 & $\delta$ & 4.0000 & 4.1027 & 0.1027 & 0.6785 & 0.4698 \\
\cline{1-7}
\multirow{9}{*}{1000} & $\omega_C$ & 3.0000 & 3.0020 & 0.0020 & 0.0298 & 0.0009 \\
 & $\phi_C$ & 0.6000 & 0.5824 & 0.0176 & 0.1051 & 0.0113 \\
 & $\kappa_C$ & 0.2500 & 0.2482 & 0.0018 & 0.0387 & 0.0015 \\
 & $\omega_Q$ & -1.8570 & -1.8543 & 0.0027 & 0.0447 & 0.0020 \\
 & $\phi_Q$ & 0.6000 & 0.5860 & 0.0140 & 0.0615 & 0.0040 \\
 & $\kappa_Q$ & 0.1700 & 0.1686 & 0.0014 & 0.0151 & 0.0002 \\
 & $\log\tau$ & 2.6450 & 2.6457 & 0.0007 & 0.0458 & 0.0021 \\
 & $\zeta$ & 1.6000 & 1.6057 & 0.0057 & 0.0765 & 0.0059 \\
 & $\delta$ & 4.0000 & 4.0160 & 0.0160 & 0.4812 & 0.2313 \\
\hline
\end{tabular}
\end{adjustbox}
\end{subtable}
\end{table}

\begin{table}[H]
\centering
\caption{Parameter estimates, ABIAS, SD and MSE for the CR-KvM model.}
\label{tab:kvm_sim}
\scriptsize
\setlength{\tabcolsep}{3pt}
\renewcommand{\arraystretch}{1.1}
\begin{subtable}[t]{0.49\textwidth}
\centering
\begin{adjustbox}{max width=\linewidth}
\begin{tabular}{|c c S[table-format=-1.4] S[table-format=-1.4]
                    S[table-format=<1.4] S[table-format=1.4] S[table-format=<2.4]|}
\hline
$T$ & Parameter & {True} & {Estimate} & {ABIAS} & {SD} & {MSE} \\
\hline
\multicolumn{7}{|c|}{I} \\
\hline
\multirow{9}{*}{100} & $\omega_C$ & 0.5000 & 0.5011 & 0.0011 & 0.0556 & 0.0031 \\
 & $\phi_C$ & 0.3000 & 0.2356 & 0.0644 & 0.2923 & 0.0893 \\
 & $\kappa_C$ & 0.4000 & 0.3776 & 0.0224 & 0.1219 & 0.0153 \\
 & $\omega_Q$ & 4.4000 & 4.5530 & 0.1530 & 0.3651 & 0.1564 \\
 & $\phi_Q$ & 0.3000 & 0.1820 & 0.1180 & 0.4362 & 0.2037 \\
 & $\kappa_Q$ & 0.2000 & 0.1831 & 0.0169 & 0.1078 & 0.0119 \\
 & $\alpha$ & 0.6931 & 0.7185 & 0.0253 & 0.0831 & 0.0075 \\
 & $\zeta$ & 2.3000 & 2.3336 & 0.0336 & 0.3134 & 0.0991 \\
 & $\delta$ & 0.5000 & 0.3144 & 0.1856 & 1.4082 & 1.9076 \\
\cline{1-7}
\multirow{9}{*}{500} & $\omega_C$ & 0.5000 & 0.5014 & 0.0014 & 0.0243 & 0.0006 \\
 & $\phi_C$ & 0.3000 & 0.2843 & 0.0157 & 0.1251 & 0.0159 \\
 & $\kappa_C$ & 0.4000 & 0.3979 & 0.0021 & 0.0481 & 0.0023 \\
 & $\omega_Q$ & 4.4000 & 4.4253 & 0.0253 & 0.1603 & 0.0263 \\
 & $\phi_Q$ & 0.3000 & 0.2507 & 0.0493 & 0.2272 & 0.0539 \\
 & $\kappa_Q$ & 0.2000 & 0.1975 & 0.0025 & 0.0463 & 0.0021 \\
 & $\alpha$ & 0.6931 & 0.6984 & 0.0053 & 0.0364 & 0.0013 \\
 & $\zeta$ & 2.3000 & 2.3087 & 0.0087 & 0.1383 & 0.0192 \\
 & $\delta$ & 0.5000 & 0.5128 & 0.0128 & 1.2354 & 1.5226 \\
\cline{1-7}
\multirow{9}{*}{1000} & $\omega_C$ & 0.5000 & 0.5025 & 0.0025 & 0.0171 & 0.0003 \\
 & $\phi_C$ & 0.3000 & 0.2918 & 0.0082 & 0.0806 & 0.0065 \\
 & $\kappa_C$ & 0.4000 & 0.4004 & 0.0004 & 0.0334 & 0.0011 \\
 & $\omega_Q$ & 4.4000 & 4.4119 & 0.0119 & 0.1028 & 0.0107 \\
 & $\phi_Q$ & 0.3000 & 0.2925 & 0.0075 & 0.1400 & 0.0196 \\
 & $\kappa_Q$ & 0.2000 & 0.1974 & 0.0026 & 0.0322 & 0.0010 \\
 & $\alpha$ & 0.6931 & 0.6949 & 0.0017 & 0.0235 & 0.0006 \\
 & $\zeta$ & 2.3000 & 2.3108 & 0.0108 & 0.0919 & 0.0085 \\
 & $\delta$ & 0.5000 & 0.5503 & 0.0503 & 0.8430 & 0.7113 \\
\hline
\multicolumn{7}{|c|}{II} \\
\hline
\multirow{9}{*}{100} & $\omega_C$ & -1.5000 & -1.4969 & 0.0031 & 0.0748 & 0.0056 \\
 & $\phi_C$ & 0.6000 & 0.4694 & 0.1306 & 0.3274 & 0.1239 \\
 & $\kappa_C$ & 0.3500 & 0.3297 & 0.0203 & 0.1169 & 0.0141 \\
 & $\omega_Q$ & 3.0000 & 3.1190 & 0.1190 & 0.3166 & 0.1142 \\
 & $\phi_Q$ & 0.6000 & 0.4802 & 0.1198 & 0.3307 & 0.1234 \\
 & $\kappa_Q$ & 0.2500 & 0.2276 & 0.0224 & 0.1044 & 0.0114 \\
 & $\alpha$ & 0.4055 & 0.4346 & 0.0292 & 0.0879 & 0.0086 \\
 & $\zeta$ & 2.0000 & 2.0197 & 0.0197 & 0.2512 & 0.0633 \\
 & $\delta$ & 2.0000 & 1.8089 & 0.1911 & 1.8443 & 3.4294 \\
\cline{1-7}
\multirow{9}{*}{500} & $\omega_C$ & -1.5000 & -1.5012 & 0.0012 & 0.0347 & 0.0012 \\
 & $\phi_C$ & 0.6000 & 0.5863 & 0.0137 & 0.0951 & 0.0092 \\
 & $\kappa_C$ & 0.3500 & 0.3458 & 0.0042 & 0.0460 & 0.0021 \\
 & $\omega_Q$ & 3.0000 & 3.0150 & 0.0150 & 0.1235 & 0.0154 \\
 & $\phi_Q$ & 0.6000 & 0.5793 & 0.0207 & 0.1268 & 0.0165 \\
 & $\kappa_Q$ & 0.2500 & 0.2474 & 0.0026 & 0.0447 & 0.0020 \\
 & $\alpha$ & 0.4055 & 0.4109 & 0.0054 & 0.0364 & 0.0014 \\
 & $\zeta$ & 2.0000 & 2.0123 & 0.0123 & 0.1040 & 0.0109 \\
 & $\delta$ & 2.0000 & 2.0370 & 0.0370 & 0.7049 & 0.4971 \\
\cline{1-7}
\multirow{9}{*}{1000} & $\omega_C$ & -1.5000 & -1.4993 & 0.0007 & 0.0244 & 0.0006 \\
 & $\phi_C$ & 0.6000 & 0.5926 & 0.0074 & 0.0664 & 0.0044 \\
 & $\kappa_C$ & 0.3500 & 0.3465 & 0.0035 & 0.0312 & 0.0010 \\
 & $\omega_Q$ & 3.0000 & 2.9993 & 0.0007 & 0.0908 & 0.0082 \\
 & $\phi_Q$ & 0.6000 & 0.5896 & 0.0104 & 0.0832 & 0.0070 \\
 & $\kappa_Q$ & 0.2500 & 0.2516 & 0.0016 & 0.0299 & 0.0009 \\
 & $\alpha$ & 0.4055 & 0.4049 & 0.0006 & 0.0263 & 0.0007 \\
 & $\zeta$ & 2.0000 & 2.0063 & 0.0063 & 0.0702 & 0.0050 \\
 & $\delta$ & 2.0000 & 2.0107 & 0.0107 & 0.4862 & 0.2359 \\
\hline
\end{tabular}
\end{adjustbox}
\end{subtable}%
\begin{subtable}[t]{0.49\textwidth}
\centering
\begin{adjustbox}{max width=\linewidth}
\begin{tabular}{|c c S[table-format=-1.4] S[table-format=-1.4]
                    S[table-format=<1.4] S[table-format=1.4] S[table-format=<2.4]|}
\hline
$T$ & Parameter & {True} & {Estimate} & {ABIAS} & {SD} & {MSE} \\
\hline
\multicolumn{7}{|c|}{III} \\
\hline
\multirow{9}{*}{100} & $\omega_C$ & 2.0000 & 2.0265 & 0.0265 & 0.3940 & 0.1556 \\
 & $\phi_C$ & 0.9500 & 0.9095 & 0.0405 & 0.0877 & 0.0093 \\
 & $\kappa_C$ & 0.5000 & 0.4950 & 0.0050 & 0.0971 & 0.0094 \\
 & $\omega_Q$ & 4.8000 & 4.9050 & 0.1050 & 0.4188 & 0.1859 \\
 & $\phi_Q$ & 0.9000 & 0.7739 & 0.1261 & 0.2944 & 0.1024 \\
 & $\kappa_Q$ & 0.1000 & 0.0820 & 0.0180 & 0.1078 & 0.0119 \\
 & $\alpha$ & 0.6931 & 0.7129 & 0.0198 & 0.0796 & 0.0067 \\
 & $\zeta$ & 2.8000 & 2.8408 & 0.0408 & 0.3221 & 0.1051 \\
 & $\delta$ & 1.0000 & 0.9102 & 0.0898 & 1.4886 & 1.9477 \\
\cline{1-7}
\multirow{9}{*}{500} & $\omega_C$ & 2.0000 & 2.0079 & 0.0079 & 0.2062 & 0.0425 \\
 & $\phi_C$ & 0.9500 & 0.9440 & 0.0060 & 0.0210 & 0.0005 \\
 & $\kappa_C$ & 0.5000 & 0.4989 & 0.0011 & 0.0418 & 0.0017 \\
 & $\omega_Q$ & 4.8000 & 4.8365 & 0.0365 & 0.1770 & 0.0326 \\
 & $\phi_Q$ & 0.9000 & 0.8587 & 0.0413 & 0.1243 & 0.0171 \\
 & $\kappa_Q$ & 0.1000 & 0.1032 & 0.0032 & 0.0330 & 0.0011 \\
 & $\alpha$ & 0.6931 & 0.6992 & 0.0060 & 0.0359 & 0.0013 \\
 & $\zeta$ & 2.8000 & 2.7989 & 0.0011 & 0.1349 & 0.0181 \\
 & $\delta$ & 1.0000 & 0.9494 & 0.0506 & 1.4528 & 1.1080 \\
\cline{1-7}
\multirow{9}{*}{1000} & $\omega_C$ & 2.0000 & 1.9917 & 0.0083 & 0.1173 & 0.0138 \\
 & $\phi_C$ & 0.9500 & 0.9464 & 0.0036 & 0.0132 & 0.0002 \\
 & $\kappa_C$ & 0.5000 & 0.4989 & 0.0011 & 0.0297 & 0.0009 \\
 & $\omega_Q$ & 4.8000 & 4.8094 & 0.0094 & 0.1285 & 0.0166 \\
 & $\phi_Q$ & 0.9000 & 0.8837 & 0.0163 & 0.0565 & 0.0035 \\
 & $\kappa_Q$ & 0.1000 & 0.0999 & 0.0001 & 0.0220 & 0.0005 \\
 & $\alpha$ & 0.6931 & 0.6951 & 0.0020 & 0.0254 & 0.0006 \\
 & $\zeta$ & 2.8000 & 2.8094 & 0.0094 & 0.0871 & 0.0077 \\
 & $\delta$ & 1.0000 & 1.0981 & 0.0981 & 0.9346 & 0.8810 \\
\hline
\multicolumn{7}{|c|}{IV} \\
\hline
\multirow{9}{*}{100} & $\omega_C$ & 3.0000 & 2.9999 & 0.0001 & 0.1083 & 0.0117 \\
 & $\phi_C$ & 0.6000 & 0.4394 & 0.1606 & 0.4058 & 0.1901 \\
 & $\kappa_C$ & 0.2500 & 0.2332 & 0.0168 & 0.1371 & 0.0190 \\
 & $\omega_Q$ & 2.8000 & 2.8633 & 0.0633 & 0.2833 & 0.0841 \\
 & $\phi_Q$ & 0.6000 & 0.4851 & 0.1149 & 0.3485 & 0.1344 \\
 & $\kappa_Q$ & 0.2500 & 0.2356 & 0.0144 & 0.1143 & 0.0132 \\
 & $\alpha$ & 0.4055 & 0.4293 & 0.0238 & 0.0883 & 0.0083 \\
 & $\zeta$ & 1.6000 & 1.6706 & 0.0706 & 0.2461 & 0.0654 \\
 & $\delta$ & 4.0000 & 4.1904 & 0.1904 & 1.6007 & 2.5922 \\
\cline{1-7}
\multirow{9}{*}{500} & $\omega_C$ & 3.0000 & 3.0027 & 0.0027 & 0.0392 & 0.0015 \\
 & $\phi_C$ & 0.6000 & 0.5669 & 0.0331 & 0.1369 & 0.0198 \\
 & $\kappa_C$ & 0.2500 & 0.2496 & 0.0004 & 0.0547 & 0.0030 \\
 & $\omega_Q$ & 2.8000 & 2.8106 & 0.0106 & 0.1297 & 0.0169 \\
 & $\phi_Q$ & 0.6000 & 0.5772 & 0.0228 & 0.1251 & 0.0161 \\
 & $\kappa_Q$ & 0.2500 & 0.2476 & 0.0024 & 0.0463 & 0.0021 \\
 & $\alpha$ & 0.4055 & 0.4064 & 0.0010 & 0.0405 & 0.0016 \\
 & $\zeta$ & 1.6000 & 1.6091 & 0.0091 & 0.1102 & 0.0122 \\
 & $\delta$ & 4.0000 & 4.0232 & 0.0232 & 0.6897 & 0.4751 \\
\cline{1-7}
\multirow{9}{*}{1000} & $\omega_C$ & 3.0000 & 3.0020 & 0.0020 & 0.0308 & 0.0010 \\
 & $\phi_C$ & 0.6000 & 0.5780 & 0.0220 & 0.1003 & 0.0105 \\
 & $\kappa_C$ & 0.2500 & 0.2521 & 0.0021 & 0.0365 & 0.0013 \\
 & $\omega_Q$ & 2.8000 & 2.8139 & 0.0139 & 0.0867 & 0.0077 \\
 & $\phi_Q$ & 0.6000 & 0.5938 & 0.0062 & 0.0801 & 0.0064 \\
 & $\kappa_Q$ & 0.2500 & 0.2500 & <0.0001 & 0.0323 & 0.0010 \\
 & $\alpha$ & 0.4055 & 0.4098 & 0.0044 & 0.0277 & 0.0008 \\
 & $\zeta$ & 1.6000 & 1.6053 & 0.0053 & 0.0748 & 0.0056 \\
 & $\delta$ & 4.0000 & 4.0267 & 0.0267 & 0.4786 & 0.2292 \\
\hline
\end{tabular}
\end{adjustbox}
\end{subtable}
\end{table}

\begin{table}[H]
\centering
\caption{Parameter estimates, ABIAS, SD and MSE of the CR-TWvM model for varying circular-linear dependence.}
\label{tab:wvm_delta}
\scriptsize
\setlength{\tabcolsep}{3pt}
\renewcommand{\arraystretch}{1.1}
\begin{subtable}[t]{0.49\textwidth}
\centering
\begin{adjustbox}{max width=\linewidth}
\begin{tabular}{|c c S[table-format=-1.4] S[table-format=-1.4]
                    S[table-format=<1.4] S[table-format=1.4] S[table-format=<1.4]|}
\hline
$T$ & Parameter & {True} & {Estimate} & {ABIAS} & {SD} & {MSE} \\
\hline
\multicolumn{7}{|c|}{A: $\delta = 0$ (independence)} \\
\hline
\multirow{9}{*}{100} & $\omega_C$ & -1.5000 & -1.4966 & 0.0034 & 0.0584 & 0.0034 \\
 & $\phi_C$ & 0.6000 & 0.5287 & 0.0713 & 0.2022 & 0.0458 \\
 & $\kappa_C$ & 0.3500 & 0.3387 & 0.0113 & 0.0927 & 0.0087 \\
 & $\omega_Q$ & -2.0230 & -2.0326 & 0.0096 & 0.1039 & 0.0108 \\
 & $\phi_Q$ & 0.6000 & 0.5528 & 0.0472 & 0.1827 & 0.0355 \\
 & $\kappa_Q$ & 0.1600 & 0.1523 & 0.0077 & 0.0416 & 0.0018 \\
 & $\log r$ & 0.4420 & 0.4585 & 0.0165 & 0.0619 & 0.0041 \\
 & $\zeta$ & 2.0000 & 2.0386 & 0.0386 & 0.1854 & 0.0357 \\
 & $\delta$ & 0.0000 & -0.0362 & 0.0362 & 1.3552 & 1.8318 \\
\cline{1-7}
\multirow{9}{*}{500} & $\omega_C$ & -1.5000 & -1.4992 & 0.0008 & 0.0318 & 0.0010 \\
 & $\phi_C$ & 0.6000 & 0.5896 & 0.0104 & 0.1062 & 0.0114 \\
 & $\kappa_C$ & 0.3500 & 0.3458 & 0.0042 & 0.0493 & 0.0024 \\
 & $\omega_Q$ & -2.0230 & -2.0190 & 0.0040 & 0.0566 & 0.0032 \\
 & $\phi_Q$ & 0.6000 & 0.5784 & 0.0216 & 0.0792 & 0.0067 \\
 & $\kappa_Q$ & 0.1600 & 0.1580 & 0.0020 & 0.0217 & 0.0005 \\
 & $\log r$ & 0.4420 & 0.4490 & 0.0070 & 0.0358 & 0.0013 \\
 & $\zeta$ & 2.0000 & 2.0099 & 0.0099 & 0.1020 & 0.0105 \\
 & $\delta$ & 0.0000 & -0.0041 & 0.0041 & 0.6408 & 0.4093 \\
\cline{1-7}
\multirow{9}{*}{1000} & $\omega_C$ & -1.5000 & -1.4958 & 0.0042 & 0.0226 & 0.0005 \\
 & $\phi_C$ & 0.6000 & 0.5836 & 0.0164 & 0.0691 & 0.0050 \\
 & $\kappa_C$ & 0.3500 & 0.3522 & 0.0022 & 0.0334 & 0.0011 \\
 & $\omega_Q$ & -2.0230 & -2.0280 & 0.0050 & 0.0416 & 0.0017 \\
 & $\phi_Q$ & 0.6000 & 0.5988 & 0.0012 & 0.0499 & 0.0025 \\
 & $\kappa_Q$ & 0.1600 & 0.1581 & 0.0019 & 0.0150 & 0.0002 \\
 & $\log r$ & 0.4420 & 0.4441 & 0.0021 & 0.0231 & 0.0005 \\
 & $\zeta$ & 2.0000 & 2.0068 & 0.0068 & 0.0672 & 0.0045 \\
 & $\delta$ & 0.0000 & 0.0163 & 0.0163 & 0.4296 & 0.1842 \\
\hline
\multicolumn{7}{|c|}{ B: $\delta = 2$} \\
\hline
\multirow{9}{*}{100} & $\omega_C$ & -1.5000 & -1.4997 & 0.0003 & 0.0630 & 0.0040 \\
 & $\phi_C$ & 0.6000 & 0.5220 & 0.0780 & 0.2246 & 0.0564 \\
 & $\kappa_C$ & 0.3500 & 0.3394 & 0.0106 & 0.0909 & 0.0083 \\
 & $\omega_Q$ & -2.0230 & -2.0163 & 0.0067 & 0.1060 & 0.0112 \\
 & $\phi_Q$ & 0.6000 & 0.5458 & 0.0542 & 0.1732 & 0.0328 \\
 & $\kappa_Q$ & 0.1600 & 0.1547 & 0.0053 & 0.0440 & 0.0020 \\
 & $\log r$ & 0.4420 & 0.4610 & 0.0190 & 0.0684 & 0.0050 \\
 & $\zeta$ & 2.0000 & 2.0113 & 0.0113 & 0.1849 & 0.0342 \\
 & $\delta$ & 2.0000 & 1.8117 & 0.1883 & 1.2648 & 1.6298 \\
\cline{1-7}
\multirow{9}{*}{500} & $\omega_C$ & -1.5000 & -1.5022 & 0.0022 & 0.0352 & 0.0012 \\
 & $\phi_C$ & 0.6000 & 0.5887 & 0.0113 & 0.0957 & 0.0093 \\
 & $\kappa_C$ & 0.3500 & 0.3456 & 0.0044 & 0.0464 & 0.0022 \\
 & $\omega_Q$ & -2.0230 & -2.0217 & 0.0013 & 0.0624 & 0.0039 \\
 & $\phi_Q$ & 0.6000 & 0.5891 & 0.0109 & 0.0806 & 0.0066 \\
 & $\kappa_Q$ & 0.1600 & 0.1563 & 0.0037 & 0.0218 & 0.0005 \\
 & $\log r$ & 0.4420 & 0.4473 & 0.0053 & 0.0326 & 0.0011 \\
 & $\zeta$ & 2.0000 & 2.0142 & 0.0142 & 0.0971 & 0.0096 \\
 & $\delta$ & 2.0000 & 2.0468 & 0.0468 & 0.6337 & 0.4024 \\
\cline{1-7}
\multirow{9}{*}{1000} & $\omega_C$ & -1.5000 & -1.4990 & 0.0010 & 0.0250 & 0.0006 \\
 & $\phi_C$ & 0.6000 & 0.5919 & 0.0081 & 0.0674 & 0.0046 \\
 & $\kappa_C$ & 0.3500 & 0.3476 & 0.0024 & 0.0309 & 0.0010 \\
 & $\omega_Q$ & -2.0230 & -2.0214 & 0.0016 & 0.0397 & 0.0016 \\
 & $\phi_Q$ & 0.6000 & 0.5947 & 0.0053 & 0.0564 & 0.0032 \\
 & $\kappa_Q$ & 0.1600 & 0.1611 & 0.0011 & 0.0148 & 0.0002 \\
 & $\log r$ & 0.4420 & 0.4408 & 0.0012 & 0.0242 & 0.0006 \\
 & $\zeta$ & 2.0000 & 2.0091 & 0.0091 & 0.0647 & 0.0043 \\
 & $\delta$ & 2.0000 & 2.0292 & 0.0292 & 0.4202 & 0.1768 \\
\hline
\end{tabular}
\end{adjustbox}
\end{subtable}%
\begin{subtable}[t]{0.49\textwidth}
\centering
\begin{adjustbox}{max width=\linewidth}
\begin{tabular}{|c c S[table-format=-1.4] S[table-format=-1.4]
                    S[table-format=<1.4] S[table-format=1.4] S[table-format=<1.4]|}
\hline
$T$ & Parameter & {True} & {Estimate} & {ABIAS} & {SD} & {MSE} \\
\hline
\multicolumn{7}{|c|}{ C: $\delta = -2$} \\
\hline
\multirow{9}{*}{100} & $\omega_C$ & -1.5000 & -1.4977 & 0.0023 & 0.0497 & 0.0025 \\
 & $\phi_C$ & 0.6000 & 0.5287 & 0.0713 & 0.1922 & 0.0419 \\
 & $\kappa_C$ & 0.3500 & 0.3497 & 0.0003 & 0.0892 & 0.0079 \\
 & $\omega_Q$ & -2.0230 & -2.0200 & 0.0030 & 0.1045 & 0.0109 \\
 & $\phi_Q$ & 0.6000 & 0.5452 & 0.0548 & 0.1679 & 0.0311 \\
 & $\kappa_Q$ & 0.1600 & 0.1563 & 0.0037 & 0.0400 & 0.0016 \\
 & $\log r$ & 0.4420 & 0.4523 & 0.0103 & 0.0667 & 0.0045 \\
 & $\zeta$ & 2.0000 & 2.0072 & 0.0072 & 0.1967 & 0.0386 \\
 & $\delta$ & -2.0000 & -2.1813 & 0.1813 & 1.3261 & 1.7855 \\
\cline{1-7}
\multirow{9}{*}{500} & $\omega_C$ & -1.5000 & -1.5006 & 0.0006 & 0.0265 & 0.0007 \\
 & $\phi_C$ & 0.6000 & 0.5819 & 0.0181 & 0.0878 & 0.0080 \\
 & $\kappa_C$ & 0.3500 & 0.3447 & 0.0053 & 0.0458 & 0.0021 \\
 & $\omega_Q$ & -2.0230 & -2.0257 & 0.0027 & 0.0568 & 0.0032 \\
 & $\phi_Q$ & 0.6000 & 0.5876 & 0.0124 & 0.0798 & 0.0065 \\
 & $\kappa_Q$ & 0.1600 & 0.1591 & 0.0009 & 0.0230 & 0.0005 \\
 & $\log r$ & 0.4420 & 0.4481 & 0.0061 & 0.0362 & 0.0013 \\
 & $\zeta$ & 2.0000 & 1.9983 & 0.0017 & 0.0992 & 0.0098 \\
 & $\delta$ & -2.0000 & -2.0814 & 0.0814 & 0.6568 & 0.4366 \\
\cline{1-7}
\multirow{9}{*}{1000} & $\omega_C$ & -1.5000 & -1.5018 & 0.0018 & 0.0202 & 0.0004 \\
 & $\phi_C$ & 0.6000 & 0.5899 & 0.0101 & 0.0655 & 0.0044 \\
 & $\kappa_C$ & 0.3500 & 0.3489 & 0.0011 & 0.0326 & 0.0011 \\
 & $\omega_Q$ & -2.0230 & -2.0212 & 0.0018 & 0.0420 & 0.0018 \\
 & $\phi_Q$ & 0.6000 & 0.5916 & 0.0084 & 0.0527 & 0.0028 \\
 & $\kappa_Q$ & 0.1600 & 0.1604 & 0.0004 & 0.0156 & 0.0002 \\
 & $\log r$ & 0.4420 & 0.4437 & 0.0017 & 0.0254 & 0.0006 \\
 & $\zeta$ & 2.0000 & 2.0011 & 0.0011 & 0.0686 & 0.0047 \\
 & $\delta$ & -2.0000 & -1.9996 & 0.0004 & 0.4293 & 0.1837 \\
\hline
\multicolumn{7}{|c|}{ D: $\delta = 5$} \\
\hline
\multirow{9}{*}{100} & $\omega_C$ & -1.5000 & -1.5022 & 0.0022 & 0.0774 & 0.0060 \\
 & $\phi_C$ & 0.6000 & 0.5306 & 0.0694 & 0.2299 & 0.0575 \\
 & $\kappa_C$ & 0.3500 & 0.3388 & 0.0112 & 0.0921 & 0.0086 \\
 & $\omega_Q$ & -2.0230 & -2.0198 & 0.0032 & 0.1212 & 0.0147 \\
 & $\phi_Q$ & 0.6000 & 0.5626 & 0.0374 & 0.1711 & 0.0306 \\
 & $\kappa_Q$ & 0.1600 & 0.1528 & 0.0072 & 0.0404 & 0.0017 \\
 & $\log r$ & 0.4420 & 0.4561 & 0.0141 & 0.0647 & 0.0044 \\
 & $\zeta$ & 2.0000 & 2.0419 & 0.0419 & 0.1934 & 0.0390 \\
 & $\delta$ & 5.0000 & 5.0687 & 0.0687 & 1.3075 & 1.7085 \\
\cline{1-7}
\multirow{9}{*}{500} & $\omega_C$ & -1.5000 & -1.4991 & 0.0009 & 0.0375 & 0.0014 \\
 & $\phi_C$ & 0.6000 & 0.5841 & 0.0159 & 0.0927 & 0.0088 \\
 & $\kappa_C$ & 0.3500 & 0.3483 & 0.0017 & 0.0482 & 0.0023 \\
 & $\omega_Q$ & -2.0230 & -2.0262 & 0.0032 & 0.0568 & 0.0032 \\
 & $\phi_Q$ & 0.6000 & 0.5958 & 0.0042 & 0.0802 & 0.0064 \\
 & $\kappa_Q$ & 0.1600 & 0.1590 & 0.0010 & 0.0238 & 0.0006 \\
 & $\log r$ & 0.4420 & 0.4435 & 0.0015 & 0.0358 & 0.0013 \\
 & $\zeta$ & 2.0000 & 2.0168 & 0.0168 & 0.1064 & 0.0116 \\
 & $\delta$ & 5.0000 & 5.1014 & 0.1014 & 0.6995 & 0.4979 \\
\cline{1-7}
\multirow{9}{*}{1000} & $\omega_C$ & -1.5000 & -1.4975 & 0.0025 & 0.0292 & 0.0009 \\
 & $\phi_C$ & 0.6000 & 0.5894 & 0.0106 & 0.0686 & 0.0048 \\
 & $\kappa_C$ & 0.3500 & 0.3524 & 0.0024 & 0.0352 & 0.0012 \\
 & $\omega_Q$ & -2.0230 & -2.0258 & 0.0028 & 0.0374 & 0.0014 \\
 & $\phi_Q$ & 0.6000 & 0.5950 & 0.0050 & 0.0521 & 0.0027 \\
 & $\kappa_Q$ & 0.1600 & 0.1578 & 0.0022 & 0.0160 & 0.0003 \\
 & $\log r$ & 0.4420 & 0.4462 & 0.0042 & 0.0251 & 0.0006 \\
 & $\zeta$ & 2.0000 & 2.0005 & 0.0005 & 0.0704 & 0.0049 \\
 & $\delta$ & 5.0000 & 4.9554 & 0.0446 & 0.4783 & 0.2300 \\
\hline
\end{tabular}
\end{adjustbox}
\end{subtable}
\end{table}

\begin{table}[H]
\centering
\caption{Parameter estimates, ABIAS, SD and MSE of the CR-BvM for varying circular-linear dependence.}
\label{tab:bvm_delta}
\scriptsize
\setlength{\tabcolsep}{3pt}
\renewcommand{\arraystretch}{1.1}
\begin{subtable}[t]{0.49\textwidth}
\centering
\begin{adjustbox}{max width=\linewidth}
\begin{tabular}{|c c S[table-format=-1.4] S[table-format=-1.4]
                    S[table-format=<1.4] S[table-format=1.4] S[table-format=<1.4]|}
\hline
$T$ & Parameter & {True} & {Estimate} & {ABIAS} & {SD} & {MSE} \\
\hline
\multicolumn{7}{|c|}{ A: $\delta = 0$ (independence)} \\
\hline
\multirow{9}{*}{100} & $\omega_C$ & -1.5000 & -1.4998 & 0.0002 & 0.0584 & 0.0034 \\
 & $\phi_C$ & 0.6000 & 0.5309 & 0.0691 & 0.2351 & 0.0599 \\
 & $\kappa_C$ & 0.3500 & 0.3342 & 0.0158 & 0.0908 & 0.0085 \\
 & $\omega_Q$ & -2.0030 & -2.0036 & 0.0006 & 0.1097 & 0.0120 \\
 & $\phi_Q$ & 0.6000 & 0.5329 & 0.0671 & 0.1868 & 0.0393 \\
 & $\kappa_Q$ & 0.1700 & 0.1613 & 0.0087 & 0.0384 & 0.0015 \\
 & $\log\tau$ & 2.7850 & 2.8130 & 0.0280 & 0.1201 & 0.0151 \\
 & $\zeta$ & 2.0000 & 2.0239 & 0.0239 & 0.2129 & 0.0457 \\
 & $\delta$ & 0.0000 & -0.0453 & 0.0453 & 1.4245 & 2.0246 \\
\cline{1-7}
\multirow{9}{*}{500} & $\omega_C$ & -1.5000 & -1.4945 & 0.0055 & 0.0307 & 0.0010 \\
 & $\phi_C$ & 0.6000 & 0.5772 & 0.0228 & 0.0977 & 0.0100 \\
 & $\kappa_C$ & 0.3500 & 0.3467 & 0.0033 & 0.0498 & 0.0025 \\
 & $\omega_Q$ & -2.0030 & -2.0059 & 0.0029 & 0.0594 & 0.0035 \\
 & $\phi_Q$ & 0.6000 & 0.5868 & 0.0132 & 0.0884 & 0.0080 \\
 & $\kappa_Q$ & 0.1700 & 0.1670 & 0.0030 & 0.0224 & 0.0005 \\
 & $\log\tau$ & 2.7850 & 2.7946 & 0.0096 & 0.0664 & 0.0045 \\
 & $\zeta$ & 2.0000 & 2.0170 & 0.0170 & 0.1192 & 0.0144 \\
 & $\delta$ & 0.0000 & 0.0036 & 0.0036 & 0.7833 & 0.6115 \\
\cline{1-7}
\multirow{9}{*}{1000} & $\omega_C$ & -1.5000 & -1.4996 & 0.0004 & 0.0201 & 0.0004 \\
 & $\phi_C$ & 0.6000 & 0.5909 & 0.0091 & 0.0665 & 0.0045 \\
 & $\kappa_C$ & 0.3500 & 0.3462 & 0.0038 & 0.0347 & 0.0012 \\
 & $\omega_Q$ & -2.0030 & -2.0068 & 0.0038 & 0.0411 & 0.0017 \\
 & $\phi_Q$ & 0.6000 & 0.5928 & 0.0072 & 0.0590 & 0.0035 \\
 & $\kappa_Q$ & 0.1700 & 0.1680 & 0.0020 & 0.0158 & 0.0003 \\
 & $\log\tau$ & 2.7850 & 2.7884 & 0.0034 & 0.0421 & 0.0018 \\
 & $\zeta$ & 2.0000 & 2.0049 & 0.0049 & 0.0672 & 0.0045 \\
 & $\delta$ & 0.0000 & 0.0006 & 0.0006 & 0.4869 & 0.2363 \\
\hline
\multicolumn{7}{|c|}{ B: $\delta = 2$} \\
\hline
\multirow{9}{*}{100} & $\omega_C$ & -1.5000 & -1.4968 & 0.0032 & 0.0618 & 0.0038 \\
 & $\phi_C$ & 0.6000 & 0.5246 & 0.0754 & 0.2026 & 0.0466 \\
 & $\kappa_C$ & 0.3500 & 0.3369 & 0.0131 & 0.0844 & 0.0073 \\
 & $\omega_Q$ & -2.0030 & -2.0058 & 0.0028 & 0.1203 & 0.0144 \\
 & $\phi_Q$ & 0.6000 & 0.5420 & 0.0580 & 0.1817 & 0.0363 \\
 & $\kappa_Q$ & 0.1700 & 0.1645 & 0.0055 & 0.0406 & 0.0017 \\
 & $\log\tau$ & 2.7850 & 2.8244 & 0.0394 & 0.1229 & 0.0166 \\
 & $\zeta$ & 2.0000 & 2.0028 & 0.0028 & 0.1998 & 0.0398 \\
 & $\delta$ & 2.0000 & 1.8310 & 0.1690 & 1.3319 & 1.7967 \\
\cline{1-7}
\multirow{9}{*}{500} & $\omega_C$ & -1.5000 & -1.5001 & 0.0001 & 0.0357 & 0.0013 \\
 & $\phi_C$ & 0.6000 & 0.5741 & 0.0259 & 0.1093 & 0.0126 \\
 & $\kappa_C$ & 0.3500 & 0.3455 & 0.0045 & 0.0505 & 0.0026 \\
 & $\omega_Q$ & -2.0030 & -2.0026 & 0.0004 & 0.0583 & 0.0034 \\
 & $\phi_Q$ & 0.6000 & 0.5729 & 0.0271 & 0.0908 & 0.0090 \\
 & $\kappa_Q$ & 0.1700 & 0.1683 & 0.0017 & 0.0219 & 0.0005 \\
 & $\log\tau$ & 2.7850 & 2.7987 & 0.0137 & 0.0645 & 0.0043 \\
 & $\zeta$ & 2.0000 & 1.9990 & 0.0010 & 0.1084 & 0.0117 \\
 & $\delta$ & 2.0000 & 1.8887 & 0.1113 & 0.7212 & 0.5308 \\
\cline{1-7}
\multirow{9}{*}{1000} & $\omega_C$ & -1.5000 & -1.4985 & 0.0015 & 0.0272 & 0.0007 \\
 & $\phi_C$ & 0.6000 & 0.5910 & 0.0090 & 0.0663 & 0.0045 \\
 & $\kappa_C$ & 0.3500 & 0.3506 & 0.0006 & 0.0365 & 0.0013 \\
 & $\omega_Q$ & -2.0030 & -2.0074 & 0.0044 & 0.0416 & 0.0017 \\
 & $\phi_Q$ & 0.6000 & 0.5852 & 0.0148 & 0.0631 & 0.0042 \\
 & $\kappa_Q$ & 0.1700 & 0.1701 & 0.0001 & 0.0160 & 0.0003 \\
 & $\log\tau$ & 2.7850 & 2.7950 & 0.0100 & 0.0468 & 0.0023 \\
 & $\zeta$ & 2.0000 & 2.0123 & 0.0123 & 0.0726 & 0.0054 \\
 & $\delta$ & 2.0000 & 2.0328 & 0.0328 & 0.4822 & 0.2328 \\
\hline
\end{tabular}
\end{adjustbox}
\end{subtable}%
\begin{subtable}[t]{0.49\textwidth}
\centering
\begin{adjustbox}{max width=\linewidth}
\begin{tabular}{|c c S[table-format=-1.4] S[table-format=-1.4]
                    S[table-format=<1.4] S[table-format=1.4] S[table-format=<1.4]|}
\hline
$T$ & Parameter & {True} & {Estimate} & {ABIAS} & {SD} & {MSE} \\
\hline
\multicolumn{7}{|c|}{ C: $\delta = -2$} \\
\hline
\multirow{9}{*}{100} & $\omega_C$ & -1.5000 & -1.4992 & 0.0008 & 0.0479 & 0.0023 \\
 & $\phi_C$ & 0.6000 & 0.5187 & 0.0813 & 0.2359 & 0.0621 \\
 & $\kappa_C$ & 0.3500 & 0.3438 & 0.0062 & 0.0880 & 0.0078 \\
 & $\omega_Q$ & -2.0030 & -2.0040 & 0.0010 & 0.1134 & 0.0128 \\
 & $\phi_Q$ & 0.6000 & 0.5391 & 0.0609 & 0.1838 & 0.0374 \\
 & $\kappa_Q$ & 0.1700 & 0.1638 & 0.0062 & 0.0380 & 0.0015 \\
 & $\log\tau$ & 2.7850 & 2.8068 & 0.0218 & 0.1247 & 0.0160 \\
 & $\zeta$ & 2.0000 & 2.0120 & 0.0120 & 0.1884 & 0.0355 \\
 & $\delta$ & -2.0000 & -2.1587 & 0.1587 & 1.3793 & 1.9214 \\
\cline{1-7}
\multirow{9}{*}{500} & $\omega_C$ & -1.5000 & -1.5019 & 0.0019 & 0.0281 & 0.0008 \\
 & $\phi_C$ & 0.6000 & 0.5804 & 0.0196 & 0.0964 & 0.0096 \\
 & $\kappa_C$ & 0.3500 & 0.3476 & 0.0024 & 0.0471 & 0.0022 \\
 & $\omega_Q$ & -2.0030 & -2.0027 & 0.0003 & 0.0590 & 0.0035 \\
 & $\phi_Q$ & 0.6000 & 0.5839 & 0.0161 & 0.0901 & 0.0083 \\
 & $\kappa_Q$ & 0.1700 & 0.1690 & 0.0010 & 0.0228 & 0.0005 \\
 & $\log\tau$ & 2.7850 & 2.7949 & 0.0099 & 0.0612 & 0.0038 \\
 & $\zeta$ & 2.0000 & 2.0021 & 0.0021 & 0.1103 & 0.0121 \\
 & $\delta$ & -2.0000 & -2.0917 & 0.0917 & 0.7474 & 0.5652 \\
\cline{1-7}
\multirow{9}{*}{1000} & $\omega_C$ & -1.5000 & -1.5011 & 0.0011 & 0.0204 & 0.0004 \\
 & $\phi_C$ & 0.6000 & 0.5941 & 0.0059 & 0.0645 & 0.0042 \\
 & $\kappa_C$ & 0.3500 & 0.3470 & 0.0030 & 0.0334 & 0.0011 \\
 & $\omega_Q$ & -2.0030 & -2.0064 & 0.0034 & 0.0434 & 0.0019 \\
 & $\phi_Q$ & 0.6000 & 0.5961 & 0.0039 & 0.0588 & 0.0035 \\
 & $\kappa_Q$ & 0.1700 & 0.1697 & 0.0003 & 0.0151 & 0.0002 \\
 & $\log\tau$ & 2.7850 & 2.7884 & 0.0034 & 0.0479 & 0.0023 \\
 & $\zeta$ & 2.0000 & 1.9971 & 0.0029 & 0.0747 & 0.0056 \\
 & $\delta$ & -2.0000 & -2.0517 & 0.0517 & 0.5357 & 0.2887 \\
\hline
\multicolumn{7}{|c|}{ D: $\delta = 5$} \\
\hline
\multirow{9}{*}{100} & $\omega_C$ & -1.5000 & -1.5040 & 0.0040 & 0.0854 & 0.0073 \\
 & $\phi_C$ & 0.6000 & 0.5477 & 0.0523 & 0.1902 & 0.0388 \\
 & $\kappa_C$ & 0.3500 & 0.3426 & 0.0074 & 0.0931 & 0.0087 \\
 & $\omega_Q$ & -2.0030 & -2.0049 & 0.0019 & 0.1074 & 0.0115 \\
 & $\phi_Q$ & 0.6000 & 0.5480 & 0.0520 & 0.1846 & 0.0367 \\
 & $\kappa_Q$ & 0.1700 & 0.1589 & 0.0111 & 0.0389 & 0.0016 \\
 & $\log\tau$ & 2.7850 & 2.8210 & 0.0360 & 0.1144 & 0.0143 \\
 & $\zeta$ & 2.0000 & 2.0568 & 0.0568 & 0.2011 & 0.0435 \\
 & $\delta$ & 5.0000 & 5.1420 & 0.1420 & 1.4633 & 2.1543 \\
\cline{1-7}
\multirow{9}{*}{500} & $\omega_C$ & -1.5000 & -1.5012 & 0.0012 & 0.0375 & 0.0014 \\
 & $\phi_C$ & 0.6000 & 0.5837 & 0.0163 & 0.1011 & 0.0105 \\
 & $\kappa_C$ & 0.3500 & 0.3471 & 0.0029 & 0.0498 & 0.0025 \\
 & $\omega_Q$ & -2.0030 & -2.0036 & 0.0006 & 0.0611 & 0.0037 \\
 & $\phi_Q$ & 0.6000 & 0.5880 & 0.0120 & 0.0892 & 0.0081 \\
 & $\kappa_Q$ & 0.1700 & 0.1696 & 0.0004 & 0.0216 & 0.0005 \\
 & $\log\tau$ & 2.7850 & 2.7934 & 0.0084 & 0.0669 & 0.0045 \\
 & $\zeta$ & 2.0000 & 2.0118 & 0.0118 & 0.1078 & 0.0117 \\
 & $\delta$ & 5.0000 & 5.0418 & 0.0418 & 0.7288 & 0.5311 \\
\cline{1-7}
\multirow{9}{*}{1000} & $\omega_C$ & -1.5000 & -1.4975 & 0.0025 & 0.0281 & 0.0008 \\
 & $\phi_C$ & 0.6000 & 0.5863 & 0.0137 & 0.0736 & 0.0056 \\
 & $\kappa_C$ & 0.3500 & 0.3478 & 0.0022 & 0.0356 & 0.0013 \\
 & $\omega_Q$ & -2.0030 & -2.0002 & 0.0028 & 0.0439 & 0.0019 \\
 & $\phi_Q$ & 0.6000 & 0.5872 & 0.0128 & 0.0583 & 0.0036 \\
 & $\kappa_Q$ & 0.1700 & 0.1704 & 0.0004 & 0.0154 & 0.0002 \\
 & $\log\tau$ & 2.7850 & 2.7856 & 0.0006 & 0.0456 & 0.0021 \\
 & $\zeta$ & 2.0000 & 2.0101 & 0.0101 & 0.0774 & 0.0061 \\
 & $\delta$ & 5.0000 & 5.0328 & 0.0328 & 0.5175 & 0.2680 \\
\hline
\end{tabular}
\end{adjustbox}
\end{subtable}
\end{table}

\begin{table}[H]
\centering
\caption{Parameter estimates, ABIAS, SD and MSE of the CR-KvM (Kumaraswamy--von Mises) model for varying circular-linear dependence.}
\label{tab:kvm_delta}
\scriptsize
\setlength{\tabcolsep}{3pt}
\renewcommand{\arraystretch}{1.1}
\begin{subtable}[t]{0.49\textwidth}
\centering
\begin{adjustbox}{max width=\linewidth}
\begin{tabular}{|c c S[table-format=-1.4] S[table-format=-1.4]
                    S[table-format=<1.4] S[table-format=1.4] S[table-format=<1.4]|}
\hline
$T$ & Parameter & {True} & {Estimate} & {ABIAS} & {SD} & {MSE} \\
\hline
\multicolumn{7}{|c|}{ A: $\delta = 0$ (independence)} \\
\hline
\multirow{9}{*}{100} & $\omega_C$ & -1.5000 & -1.4967 & 0.0033 & 0.0586 & 0.0034 \\
 & $\phi_C$ & 0.6000 & 0.5283 & 0.0717 & 0.2015 & 0.0456 \\
 & $\kappa_C$ & 0.3500 & 0.3389 & 0.0111 & 0.0925 & 0.0087 \\
 & $\omega_Q$ & 3.0000 & 3.0768 & 0.0768 & 0.2247 & 0.0562 \\
 & $\phi_Q$ & 0.6000 & 0.5157 & 0.0843 & 0.2794 & 0.0849 \\
 & $\kappa_Q$ & 0.2500 & 0.2416 & 0.0084 & 0.0872 & 0.0076 \\
 & $\alpha$ & 0.4055 & 0.4233 & 0.0179 & 0.0681 & 0.0049 \\
 & $\zeta$ & 2.0000 & 2.0417 & 0.0417 & 0.1984 & 0.0410 \\
 & $\delta$ & 0.0000 & -0.0138 & 0.0138 & 1.5169 & 2.2934 \\
\cline{1-7}
\multirow{9}{*}{500} & $\omega_C$ & -1.5000 & -1.4992 & 0.0008 & 0.0318 & 0.0010 \\
 & $\phi_C$ & 0.6000 & 0.5897 & 0.0103 & 0.1060 & 0.0113 \\
 & $\kappa_C$ & 0.3500 & 0.3457 & 0.0043 & 0.0493 & 0.0024 \\
 & $\omega_Q$ & 3.0000 & 3.0213 & 0.0213 & 0.1284 & 0.0169 \\
 & $\phi_Q$ & 0.6000 & 0.5629 & 0.0371 & 0.1219 & 0.0162 \\
 & $\kappa_Q$ & 0.2500 & 0.2488 & 0.0012 & 0.0440 & 0.0019 \\
 & $\alpha$ & 0.4055 & 0.4132 & 0.0078 & 0.0387 & 0.0016 \\
 & $\zeta$ & 2.0000 & 2.0115 & 0.0115 & 0.1108 & 0.0124 \\
 & $\delta$ & 0.0000 & 0.0087 & 0.0087 & 0.7392 & 0.5447 \\
\cline{1-7}
\multirow{9}{*}{1000} & $\omega_C$ & -1.5000 & -1.4958 & 0.0042 & 0.0226 & 0.0005 \\
 & $\phi_C$ & 0.6000 & 0.5836 & 0.0164 & 0.0690 & 0.0050 \\
 & $\kappa_C$ & 0.3500 & 0.3521 & 0.0021 & 0.0334 & 0.0011 \\
 & $\omega_Q$ & 3.0000 & 3.0152 & 0.0152 & 0.0855 & 0.0075 \\
 & $\phi_Q$ & 0.6000 & 0.5967 & 0.0033 & 0.0750 & 0.0056 \\
 & $\kappa_Q$ & 0.2500 & 0.2470 & 0.0030 & 0.0310 & 0.0010 \\
 & $\alpha$ & 0.4055 & 0.4079 & 0.0025 & 0.0255 & 0.0007 \\
 & $\zeta$ & 2.0000 & 2.0081 & 0.0081 & 0.0723 & 0.0053 \\
 & $\delta$ & 0.0000 & 0.0281 & 0.0281 & 0.4954 & 0.2454 \\
\hline
\multicolumn{7}{|c|}{ B: $\delta = 2$} \\
\hline
\multirow{9}{*}{100} & $\omega_C$ & -1.5000 & -1.4991 & 0.0009 & 0.0626 & 0.0039 \\
 & $\phi_C$ & 0.6000 & 0.5170 & 0.0830 & 0.2343 & 0.0616 \\
 & $\kappa_C$ & 0.3500 & 0.3395 & 0.0105 & 0.0894 & 0.0081 \\
 & $\omega_Q$ & 3.0000 & 3.0671 & 0.0671 & 0.2629 & 0.0734 \\
 & $\phi_Q$ & 0.6000 & 0.5106 & 0.0894 & 0.2520 & 0.0713 \\
 & $\kappa_Q$ & 0.2500 & 0.2461 & 0.0039 & 0.0851 & 0.0072 \\
 & $\alpha$ & 0.4055 & 0.4259 & 0.0205 & 0.0745 & 0.0060 \\
 & $\zeta$ & 2.0000 & 2.0124 & 0.0124 & 0.1998 & 0.0400 \\
 & $\delta$ & 2.0000 & 1.8409 & 0.1591 & 1.4267 & 2.0540 \\
\cline{1-7}
\multirow{9}{*}{500} & $\omega_C$ & -1.5000 & -1.5021 & 0.0021 & 0.0350 & 0.0012 \\
 & $\phi_C$ & 0.6000 & 0.5874 & 0.0126 & 0.0962 & 0.0094 \\
 & $\kappa_C$ & 0.3500 & 0.3454 & 0.0046 & 0.0466 & 0.0022 \\
 & $\omega_Q$ & 3.0000 & 3.0182 & 0.0182 & 0.1249 & 0.0159 \\
 & $\phi_Q$ & 0.6000 & 0.5815 & 0.0185 & 0.1259 & 0.0161 \\
 & $\kappa_Q$ & 0.2500 & 0.2448 & 0.0052 & 0.0429 & 0.0019 \\
 & $\alpha$ & 0.4055 & 0.4115 & 0.0061 & 0.0356 & 0.0013 \\
 & $\zeta$ & 2.0000 & 2.0165 & 0.0165 & 0.1062 & 0.0115 \\
 & $\delta$ & 2.0000 & 2.0582 & 0.0582 & 0.7214 & 0.5220 \\
\cline{1-7}
\multirow{9}{*}{1000} & $\omega_C$ & -1.5000 & -1.4990 & 0.0010 & 0.0250 & 0.0006 \\
 & $\phi_C$ & 0.6000 & 0.5917 & 0.0083 & 0.0679 & 0.0047 \\
 & $\kappa_C$ & 0.3500 & 0.3473 & 0.0027 & 0.0310 & 0.0010 \\
 & $\omega_Q$ & 3.0000 & 2.9942 & 0.0058 & 0.0905 & 0.0082 \\
 & $\phi_Q$ & 0.6000 & 0.5894 & 0.0106 & 0.0832 & 0.0070 \\
 & $\kappa_Q$ & 0.2500 & 0.2515 & 0.0015 & 0.0311 & 0.0010 \\
 & $\alpha$ & 0.4055 & 0.4039 & 0.0016 & 0.0267 & 0.0007 \\
 & $\zeta$ & 2.0000 & 2.0073 & 0.0073 & 0.0695 & 0.0049 \\
 & $\delta$ & 2.0000 & 2.0180 & 0.0180 & 0.4876 & 0.2372 \\
\hline
\end{tabular}
\end{adjustbox}
\end{subtable}%
\begin{subtable}[t]{0.49\textwidth}
\centering
\begin{adjustbox}{max width=\linewidth}
\begin{tabular}{|c c S[table-format=-1.4] S[table-format=-1.4]
                    S[table-format=<1.4] S[table-format=1.4] S[table-format=<1.4]|}
\hline
$T$ & Parameter & {True} & {Estimate} & {ABIAS} & {SD} & {MSE} \\
\hline
\multicolumn{7}{|c|}{ C: $\delta = -2$} \\
\hline
\multirow{9}{*}{100} & $\omega_C$ & -1.5000 & -1.4982 & 0.0018 & 0.0492 & 0.0024 \\
 & $\phi_C$ & 0.6000 & 0.5287 & 0.0713 & 0.1933 & 0.0423 \\
 & $\kappa_C$ & 0.3500 & 0.3500 & <0.0001 & 0.0903 & 0.0081 \\
 & $\omega_Q$ & 3.0000 & 3.0373 & 0.0373 & 0.2424 & 0.0599 \\
 & $\phi_Q$ & 0.6000 & 0.5139 & 0.0861 & 0.2546 & 0.0720 \\
 & $\kappa_Q$ & 0.2500 & 0.2440 & 0.0060 & 0.0829 & 0.0069 \\
 & $\alpha$ & 0.4055 & 0.4155 & 0.0100 & 0.0725 & 0.0053 \\
 & $\zeta$ & 2.0000 & 2.0106 & 0.0106 & 0.2073 & 0.0429 \\
 & $\delta$ & -2.0000 & -2.1561 & 0.1561 & 1.4402 & 2.0917 \\
\cline{1-7}
\multirow{9}{*}{500} & $\omega_C$ & -1.5000 & -1.5008 & 0.0008 & 0.0266 & 0.0007 \\
 & $\phi_C$ & 0.6000 & 0.5815 & 0.0185 & 0.0892 & 0.0083 \\
 & $\kappa_C$ & 0.3500 & 0.3445 & 0.0055 & 0.0458 & 0.0021 \\
 & $\omega_Q$ & 3.0000 & 3.0271 & 0.0271 & 0.1292 & 0.0174 \\
 & $\phi_Q$ & 0.6000 & 0.5748 & 0.0252 & 0.1233 & 0.0158 \\
 & $\kappa_Q$ & 0.2500 & 0.2509 & 0.0009 & 0.0458 & 0.0021 \\
 & $\alpha$ & 0.4055 & 0.4123 & 0.0068 & 0.0399 & 0.0016 \\
 & $\zeta$ & 2.0000 & 2.0010 & 0.0010 & 0.1091 & 0.0119 \\
 & $\delta$ & -2.0000 & -2.0638 & 0.0638 & 0.7615 & 0.5819 \\
\cline{1-7}
\multirow{9}{*}{1000} & $\omega_C$ & -1.5000 & -1.5019 & 0.0019 & 0.0205 & 0.0004 \\
 & $\phi_C$ & 0.6000 & 0.5895 & 0.0105 & 0.0653 & 0.0044 \\
 & $\kappa_C$ & 0.3500 & 0.3487 & 0.0013 & 0.0326 & 0.0011 \\
 & $\omega_Q$ & 3.0000 & 3.0043 & 0.0043 & 0.0934 & 0.0087 \\
 & $\phi_Q$ & 0.6000 & 0.5855 & 0.0145 & 0.0794 & 0.0065 \\
 & $\kappa_Q$ & 0.2500 & 0.2512 & 0.0012 & 0.0295 & 0.0009 \\
 & $\alpha$ & 0.4055 & 0.4073 & 0.0018 & 0.0275 & 0.0008 \\
 & $\zeta$ & 2.0000 & 2.0034 & 0.0034 & 0.0740 & 0.0055 \\
 & $\delta$ & -2.0000 & -1.9854 & 0.0146 & 0.4943 & 0.2437 \\
\hline
\multicolumn{7}{|c|}{ D: $\delta = 5$} \\
\hline
\multirow{9}{*}{100} & $\omega_C$ & -1.5000 & -1.5033 & 0.0033 & 0.0776 & 0.0060 \\
 & $\phi_C$ & 0.6000 & 0.5348 & 0.0652 & 0.2240 & 0.0543 \\
 & $\kappa_C$ & 0.3500 & 0.3366 & 0.0134 & 0.0912 & 0.0085 \\
 & $\omega_Q$ & 3.0000 & 3.0484 & 0.0484 & 0.2431 & 0.0612 \\
 & $\phi_Q$ & 0.6000 & 0.5225 & 0.0775 & 0.2804 & 0.0844 \\
 & $\kappa_Q$ & 0.2500 & 0.2396 & 0.0104 & 0.0865 & 0.0076 \\
 & $\alpha$ & 0.4055 & 0.4190 & 0.0135 & 0.0707 & 0.0052 \\
 & $\zeta$ & 2.0000 & 2.0368 & 0.0368 & 0.2009 & 0.0416 \\
 & $\delta$ & 5.0000 & 5.0567 & 0.0567 & 1.3900 & 1.9290 \\
\cline{1-7}
\multirow{9}{*}{500} & $\omega_C$ & -1.5000 & -1.5001 & 0.0001 & 0.0370 & 0.0014 \\
 & $\phi_C$ & 0.6000 & 0.5824 & 0.0176 & 0.0920 & 0.0087 \\
 & $\kappa_C$ & 0.3500 & 0.3481 & 0.0019 & 0.0489 & 0.0024 \\
 & $\omega_Q$ & 3.0000 & 3.0109 & 0.0109 & 0.1317 & 0.0174 \\
 & $\phi_Q$ & 0.6000 & 0.5828 & 0.0172 & 0.1350 & 0.0185 \\
 & $\kappa_Q$ & 0.2500 & 0.2476 & 0.0024 & 0.0473 & 0.0022 \\
 & $\alpha$ & 0.4055 & 0.4065 & 0.0011 & 0.0386 & 0.0015 \\
 & $\zeta$ & 2.0000 & 2.0160 & 0.0160 & 0.1092 & 0.0121 \\
 & $\delta$ & 5.0000 & 5.0917 & 0.0917 & 0.7355 & 0.5476 \\
\cline{1-7}
\multirow{9}{*}{1000} & $\omega_C$ & -1.5000 & -1.4971 & 0.0029 & 0.0284 & 0.0008 \\
 & $\phi_C$ & 0.6000 & 0.5882 & 0.0118 & 0.0733 & 0.0055 \\
 & $\kappa_C$ & 0.3500 & 0.3537 & 0.0037 & 0.0335 & 0.0011 \\
 & $\omega_Q$ & 3.0000 & 3.0185 & 0.0185 & 0.0918 & 0.0087 \\
 & $\phi_Q$ & 0.6000 & 0.5909 & 0.0091 & 0.0779 & 0.0061 \\
 & $\kappa_Q$ & 0.2500 & 0.2476 & 0.0024 & 0.0317 & 0.0010 \\
 & $\alpha$ & 0.4055 & 0.4092 & 0.0038 & 0.0282 & 0.0008 \\
 & $\zeta$ & 2.0000 & 2.0020 & 0.0020 & 0.0744 & 0.0055 \\
 & $\delta$ & 5.0000 & 4.9696 & 0.0304 & 0.5179 & 0.2682 \\
\hline
\end{tabular}
\end{adjustbox}
\end{subtable}
\end{table}

\section{Synthetic data analysis}\label{sec:synthetic}
\begin{figure}[ht]
    \centering
    \includegraphics[width=\linewidth, height=8cm]{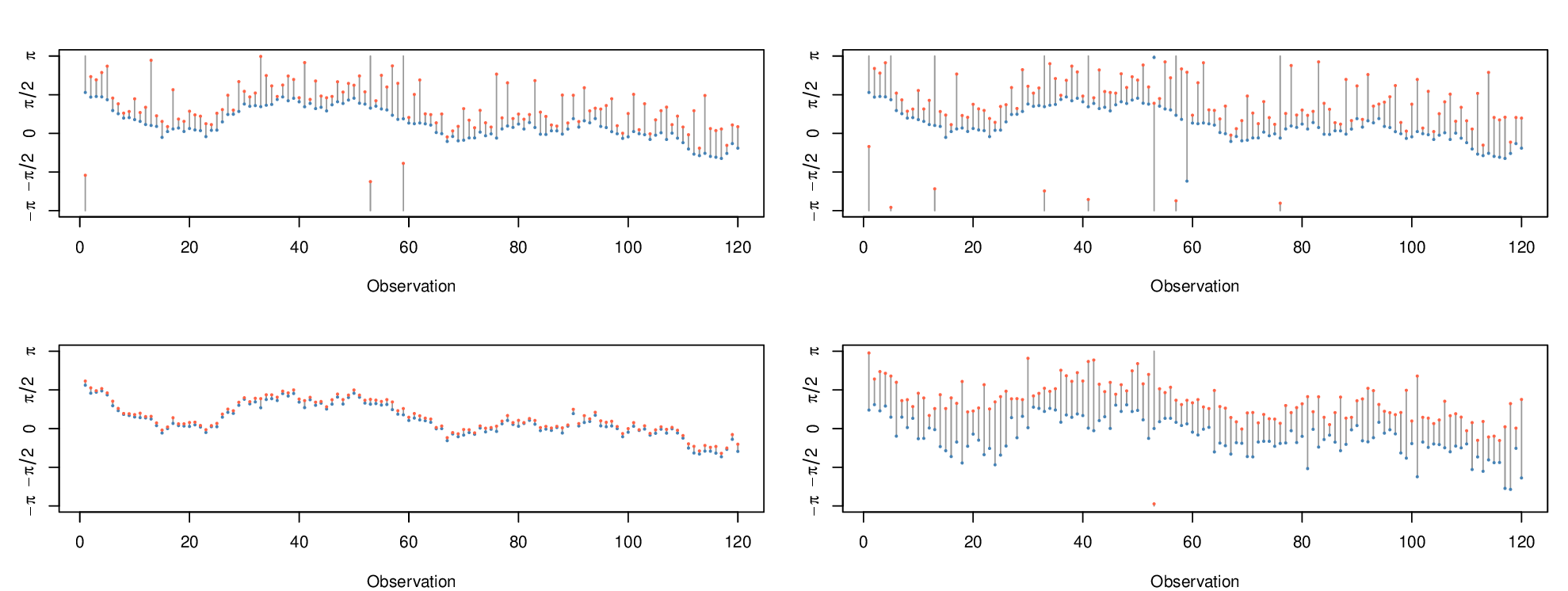}
    \caption{Synthetic data scenarios, A and B (top), C and D (bottom).}
    \label{synthetic}
\end{figure}

To investigate the performance of the proposed centre--range approaches (CR-TWvM, CR-BvM and CR-KvM) as an alternative to the min--max approach (IN-DCS) of \cite{mohan2026interval}, in this section, a synthetic data analysis is performed. To perform this analysis, we simulate different scenarios and fit both models.
Four scenarios are considered to illustrate different patterns of circular intervals. In Scenarios A and B, the lower endpoints follow a clear temporal pattern, while the upper endpoints show greater variability, with some observations crossing the circular cut. Scenario C contains narrow intervals whose endpoints move closely together over time. Scenario D contains wider intervals, with substantial variation in both endpoints. The plots also include intervals that cross the circular cut, appearing near both $-\pi$ and $\pi$.
The plots of the first $n=120$ observations of the synthetic data are provided in Figure \ref{synthetic}.
For each scenario, samples of size $1000$ are generated and repeated $1000$ times. The IN-DCS model, CR-TWvM, CR-BvM, and CR-KvM are fitted to the synthetic data observations. The dependence parameter from the IN-DCS model is $\lambda$. We use the interval circular mean error (IN-CME) in \cite{mohan2026cinar}, which is defined as
\begin{equation}
\text{IN-CME}(\widehat{\Theta},\Theta) = \frac{1}{T-p}\sum_{t=1}^{T}\left[d\!\left(\theta_t^{L},\widehat{\theta}_t^{L}\right) + d\!\left(\theta_t^{U},\widehat{\theta}_t^{U}\right)\right], ~ d(\theta,\widehat{\theta}) = 1 - \cos(\theta - \widehat{\theta}),
\label{eq:incmse}
\end{equation}
where $p, T$ are the number of parameters and observations, respectively.
The average Bayesian Information Criterion (BIC) and IN-CME are reported in Table~\ref{tab:four_scenarios}. The smallest BIC and IN-CME values have been bolded.
\begin{table}[H]
\centering
\caption{BIC and IN-CME results for the synthetic data.}
\label{tab:four_scenarios}
\renewcommand{\arraystretch}{1.1}
\small
\begin{adjustbox}{max width=\linewidth}
\begin{tabular}{|l|l|r|c|}
\hline
Scenario & Model & \multicolumn{1}{c|}{BIC} & IN-CME \\
\hline
\multirow{8}{*}{\shortstack[l]{A}}
 & IN-DCS ($\lambda=0$)      & \textbf{816.1455} & \textbf{0.2084} \\
 & IN-DCS                    & 817.4864  & 0.2110 \\
 & CR-TWvM                    & 1102.4486 & 0.2154 \\
 & CR-BvM                    & 1098.0207 & 0.2193 \\
 & CR-KvM                    & 1119.2674 & 0.2165 \\
 & CR-TWvM ($\delta = 0$)     & 1204.0607 & 0.2100 \\
 & CR-BvM ($\delta = 0$)     & 1199.6329 & 0.2140 \\
 & CR-KvM ($\delta = 0$)     & 1220.8796 & 0.2111 \\
\hline
\multirow{8}{*}{\shortstack[l]{B}}
 & IN-DCS ($\lambda=0$)      & \textbf{1284.2300} & \textbf{0.3025} \\
 & IN-DCS                    & 1289.2243 & 0.3033 \\
 & CR-TWvM                    & 1579.6737 & 0.3225 \\
 & CR-BvM                    & 1580.5169 & 0.3216 \\
 & CR-KvM                    & 1596.8272 & 0.3240 \\
 & CR-TWvM ($\delta = 0$)     & 1686.5172 & 0.3104 \\
 & CR-BvM ($\delta = 0$)     & 1687.3603 & 0.3096 \\
 & CR-KvM ($\delta = 0$)     & 1703.6706 & 0.3120 \\
\hline
\multirow{8}{*}{\shortstack[l]{C}}
 & IN-DCS ($\lambda=0$)      & $-231.2915$  & 0.0447 \\
 & IN-DCS                    & $-485.6095$  & 0.0515 \\
 & CR-TWvM                    & $-1564.7503$ & 0.0441 \\
 & CR-BvM                    & $-1543.2432$ & 0.0691 \\
 & CR-KvM                    & $-1564.6160$ & 0.0440 \\
 & CR-TWvM ($\delta = 0$)     & $\mathbf{-1568.1696}$ & $\mathbf{0.0441} $\\
 & CR-BvM ($\delta = 0$)     & $-1546.6625$ & 0.0691 \\
 & CR-KvM ($\delta = 0$)     & $\mathbf{-1568.0353}$ & \textbf{0.0440} \\
\hline
\multirow{8}{*}{\shortstack[l]{D}}
 & IN-DCS ($\lambda=0$)      & 1394.2322 & 0.2140 \\
 & IN-DCS                    & 1224.4516 & 0.1943 \\
 & CR-TWvM                    & 1028.4272 & \textbf{0.1773} \\
 & CR-BvM                    & \textbf{1013.5038} & 0.1778 \\
 & CR-KvM                    & 1047.2082 & 0.1794 \\
 & CR-TWvM ($\delta = 0$)     & 1103.3321 & 0.1773 \\
 & CR-BvM ($\delta = 0$)     & 1088.4087 & 0.1779 \\
 & CR-KvM ($\delta = 0$)     & 1122.1132 & 0.1794 \\
\hline
\end{tabular}
\end{adjustbox}
\end{table}

The results from Table~\ref{tab:four_scenarios} indicate that in Scenarios~A and~B,
where the variation within each interval is one-sided, the IN-DCS models are preferred with the smallest BIC and IN-CME.
In Scenarios~C, where intervals are narrow on a small arc, all CR methods are preferred. Similarly, in Scenario D for wide arcs, the same conclusion is drawn. These findings suggest that model performance depends on the pattern of endpoint variation and interval width. In the next section, an application will be discussed.

\section{Application} \label{sec:application} 
\begin{figure}[ht]
    \centering
        \includegraphics[width=\linewidth]{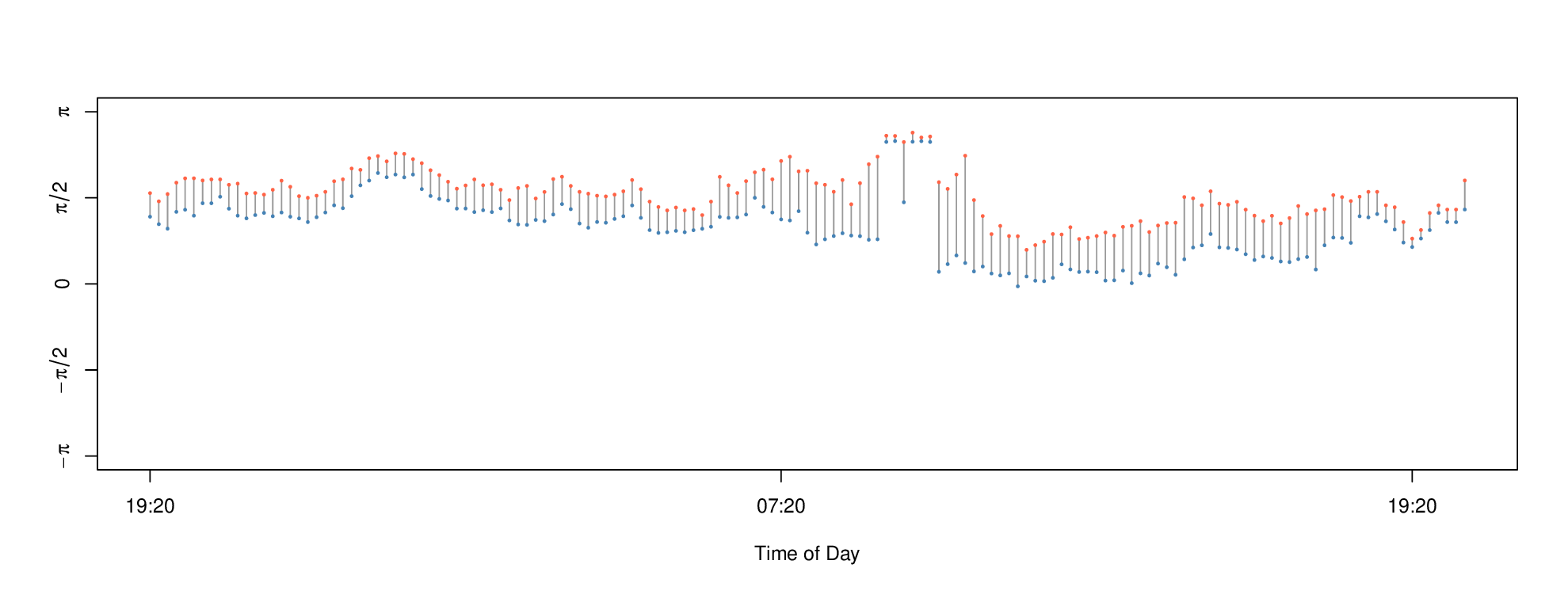}
        \includegraphics[width=\linewidth]{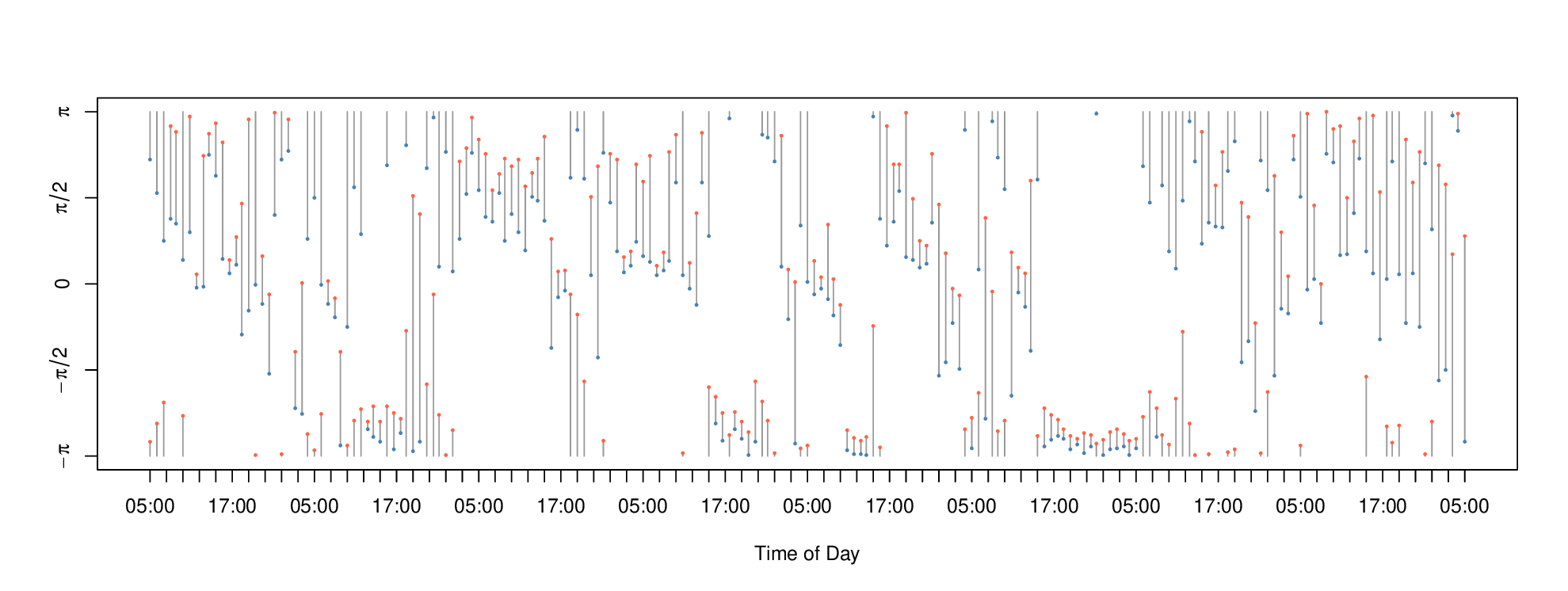}
    \caption{Interval-valued wind direction data for South Africa (top) and Brazil (bottom).}
    \label{fig:applications}
\end{figure}

The practical utility of the CR-TWVM, CR-BvM and CR-KvM models is illustrated using interval-valued wind direction data
recorded at two locations: a South African wind farm, where the wind direction was recorded every
10 minutes and summarised into hourly intervals, and a Brazilian meteorological station, where the
wind direction was recorded every hour and summarised into 5-hourly intervals. The two series are
plotted in Figure~\ref{fig:applications}, with $151$ observations and $231$ observations for the South African and Brazil dataset, respectively. The South African intervals are relatively narrow, whereas
the Brazilian intervals are wider and more variable, so that the two datasets represent different
situations that arise in practice.

The CR-TWvM, CR-BvM and CR-KvM models are compared with the independent and dependent IN-DCS models. For the
South African data, the parameter estimates of the IN-DCS models are given in
Table~\ref{tab:indcs_estimates_SA} and those of the CR-TWVM, CR-BvM and CR-KvM  models in Table~\ref{tab:cr_estimates_SA}. A likelihood ratio test of $H_0\!: \lambda = 0$ and $H_0\!: \delta = 0$, which test the circular--circular and circular--linear dependence, respectively has also been performed.  The test statistic (LRT) and p-value (p) is reported in Tables \ref{tab:indcs_estimates_SA} \ref{tab:cr_estimates_SA}, \ref{tab:indcs_estimates_BR} and \ref{tab:cr_estimates_BR}. The models are compared in Table~\ref{tab:comparison_SA} using the BIC and the
IN-CME. Since the log-likelihoods of the CR-TWvM, CR-BvM and CR-KvM  models use  the centre and the bounded range, they are expressed on the scale of the bounds by subtracting
$T\log(2\pi)$ before their BIC values are compared with those of the IN-DCS models. The corresponding
results for the Brazilian data are given in Tables~\ref{tab:indcs_estimates_BR},
\ref{tab:cr_estimates_BR} and~\ref{tab:comparison_BR}. The lowest BIC has been bolded.

\begin{table}[H]
\centering
\caption{Parameter estimates of the IN-DCS models for the South African data.}
\label{tab:indcs_estimates_SA}
\renewcommand{\arraystretch}{1.1}
\small
\begin{tabular}{|l|c|c|}
\hline
Parameter & Independent & Dependent \\
\hline
$\kappa_L$   & 1.0235  & 0.9866 \\
$\phi_L$     & 0.8481  & 0.8635 \\
$\mu_L$      & 1.0790  & 1.0747 \\
$\upsilon_L$ & 12.5832 & 12.9223 \\
$\kappa_U$   & 0.9996  & 0.9810 \\
$\phi_U$     & 0.8938  & 0.9047 \\
$\mu_U$      & 1.6546  & 1.6421 \\
$\upsilon_U$ & 28.8473 & 29.5690 \\
$\lambda$    & - & 2.5489 \\
\hline
\multicolumn{3}{|l|}{LR test of $H_0\!: \lambda = 0$: $\text{LRT }= 5.703$, $p = 0.0169$} \\
\hline
\end{tabular}
\end{table}

\begin{table}[H]
\centering
\caption{Parameter estimates of the CR-TWVM, CR-BvM and CR-KvM models for the South African data. }
\label{tab:cr_estimates_SA}
\renewcommand{\arraystretch}{1.1}
\small
\begin{tabular}{|l|c|c|c|}
\hline
Parameter & CR-TWvM & CR-BvM & CR-KvM \\
\hline
\multicolumn{4}{|c|}{{Centre, dependent model}} \\
\hline
$\mu_C$    & \multicolumn{3}{c|}{1.4880} \\
$\phi_C$   & \multicolumn{3}{c|}{0.9427} \\
$\kappa_C$ & \multicolumn{3}{c|}{1.0728} \\
$\zeta$    & \multicolumn{3}{c|}{3.7131} \\
$\delta$   & \multicolumn{3}{c|}{4.7051} \\
\hline
\multicolumn{4}{|c|}{{Centre, independent model}} \\
\hline
$\mu_C$    & \multicolumn{3}{c|}{1.3884} \\
$\phi_C$   & \multicolumn{3}{c|}{0.8871} \\
$\kappa_C$ & \multicolumn{3}{c|}{1.0863} \\
$\zeta$    & \multicolumn{3}{c|}{3.1664} \\
\hline
\multicolumn{4}{|c|}{{Range}} \\
\hline
$\omega_Q$ & $-2.3931$ & $-2.2492$ & 7.4180 \\
$\phi_Q$   & 0.9681 & 0.5565 & 0.9681 \\
$\kappa_Q$ & 0.0265 & 0.0988 & 0.2552 \\
Shape      & $\log r$: 1.1320 & $\log\tau$: 4.0350 & $\alpha$: 1.1314 \\
\hline
\multicolumn{4}{|l|}{LR test of $H_0\!: \delta = 0$: $\text{LRT }= 12.520$, $p = 0.0004$} \\
\hline
\end{tabular}
\end{table}
\begin{table}[H]
\centering
\caption{BIC and IN-CME of the IN-DCS and CR-TWVM, CR-BvM and CR-KvM  models for the South African data
.}
\label{tab:comparison_SA}
\renewcommand{\arraystretch}{1.1}
\small
\begin{tabular}{|l|c|c|c|c|}
\hline
 & \multicolumn{2}{c|}{BIC} & \multicolumn{2}{c|}{IN-CME} \\
\cline{2-5}
Model & Independent & Dependent & Independent & Dependent \\
\hline
IN-DCS & 16.220    & 15.534    & 0.0581 & 0.0593 \\
CR-TWvM & $-50.590$ & $\mathbf{-58.092}$ & 0.0549 & 0.0555 \\
CR-BvM & $-22.909$ & $-30.411$ & 0.0543 & 0.0546 \\
CR-KvM & $-50.495$ & $-57.997$ & 0.0550 & 0.0555 \\
\hline
\end{tabular}
\end{table}

\begin{table}[H]
\centering
\caption{Parameter estimates of the IN-DCS models for the Brazilian data ($T = 231$).}
\label{tab:indcs_estimates_BR}
\renewcommand{\arraystretch}{1.1}
\small
\begin{tabular}{|l|c|c|}
\hline
Parameter & Independent & Dependent \\
\hline
$\kappa_L$   & 1.3141 & 1.2953 \\
$\phi_L$     & 0.9136 & 0.8985 \\
$\mu_L$      & 0.4169 & 0.3523 \\
$\upsilon_L$ & 0.7754 & 0.8161 \\
$\kappa_U$   & 0.9200 & 1.0310 \\
$\phi_U$     & 0.3404 & 0.3222 \\
$\mu_U$      & 2.7003 & 2.7621 \\
$\upsilon_U$ & 1.1126 & 1.1086 \\
$\lambda$    & - & $-0.1958$ \\
\hline
\multicolumn{3}{|l|}{LR test of $H_0\!: \lambda = 0$: $\text{LRT }= 6.421$, $p = 0.0113$} \\
\hline
\end{tabular}
\end{table}
\begin{table}[H]
\centering
\caption{Parameter estimates of the CR-TWVM, CR-BvM and CR-KvM models for the Brazilian data .}
\label{tab:cr_estimates_BR}
\renewcommand{\arraystretch}{1.1}
\small
\begin{tabular}{|l|c|c|c|}
\hline
Parameter & CR-TWvM & CR-BvM & CR-KvM \\
\hline
\multicolumn{4}{|c|}{{Centre, dependent model}} \\
\hline
$\mu_C$    & \multicolumn{3}{c|}{2.0314} \\
$\phi_C$   & \multicolumn{3}{c|}{0.3396} \\
$\kappa_C$ & \multicolumn{3}{c|}{1.0795} \\
$\zeta$    & \multicolumn{3}{c|}{0.4946} \\
$\delta$   & \multicolumn{3}{c|}{1.1909} \\
\hline
\multicolumn{4}{|c|}{{Centre, independent model}} \\
\hline
$\mu_C$    & \multicolumn{3}{c|}{1.7043} \\
$\phi_C$   & \multicolumn{3}{c|}{0.4699} \\
$\kappa_C$ & \multicolumn{3}{c|}{1.1451} \\
$\zeta$    & \multicolumn{3}{c|}{0.1886} \\
\hline
\multicolumn{4}{|c|}{{Range}} \\
\hline
$\omega_Q$ & $-1.2777$ & $-1.0817$ & 1.4035 \\
$\phi_Q$   & 0.7518 & 0.3945 & 0.7342 \\
$\kappa_Q$ & 0.1700 & 0.1942 & 0.2398 \\
Shape      & $\log r$: 0.2877 & $\log\tau$: 1.6796 & $\alpha$: 0.1938 \\
\hline
\multicolumn{4}{|l|}{LR test of $H_0\!: \delta = 0$: $\text{LRT }= 6.411$, $p = 0.0113$} \\
\hline
\end{tabular}
\end{table}
\begin{table}[H]
\centering
\caption{BIC and IN-CME of the IN-DCS and CR-TWVM, CR-BvM and CR-KvM  models for the Brazilian data
.}
\label{tab:comparison_BR}
\renewcommand{\arraystretch}{1.1}
\small
\begin{tabular}{|l|c|c|c|c|}
\hline
 & \multicolumn{2}{c|}{BIC} & \multicolumn{2}{c|}{IN-CME} \\
\cline{2-5}
Model & Independent & Dependent & Independent & Dependent \\
\hline
IN-DCS & 1563.343 & 1562.364 & 1.1540 & 1.1394 \\
CR-TWvM & 1387.706 & 1386.737 & 1.0849 & 1.0781 \\
CR-BvM & 1379.937 & 1378.969 & 1.0856 & \textbf{1.0781} \\
CR-KvM & 1382.479 & 1381.511 & 1.0864 & 1.0800 \\
\hline
\end{tabular}
\end{table}

For the South African data, the dependence between the bounds is significant in the IN-DCS models as provided in Table~\ref{tab:indcs_estimates_SA}. The dependent model has the lower BIC as compared to the independent IN-DCS model. For the CR-TWVM, CR-BvM and CR-KvM  models, the hypothesis $\delta = 0$ is rejected as well as can be seen in Table~\ref{tab:cr_estimates_SA}. From Table~\ref{tab:comparison_SA}, the CR-TWVM, CR-BvM and CR-KvM  models are preferred to the IN-DCS models on both performance metrics since the lowest BIC and IN-CME is for the
CR-TWVM, CR-BvM and CR-KvM models. Considering only the proposed approaches in this paper, the lowest BIC is for the dependent models.

The Brazilian data intervals are considerably wider and more variable, Both tests for dependency are
significant as provided in
Table~\ref{tab:indcs_estimates_BR} and Table~\ref{tab:cr_estimates_BR}. The advantage of the CR-TWVM, CR-BvM and CR-KvM models is
more evident in this dataset  since the best CR-TWVM, CR-BvM and CR-KvM 
models has lower BIC values.

For both datasets, the centre--range representation therefore provides a better description of the
intervals. This is
consistent with the results of Section~\ref{sec:synthetic}, since in both datasets the intervals follow a similar shape when the
CR-TWVM, CR-BvM and CR-KvM models were preferred in the synthetic data analysis. The differences in performance evaluation between the fit of three models for the ranges are small however it is noted that the truncated Weibull distribution has the lowest BIC for the South African data and the beta distribution for the Brazilian data.
In conclusion, the applications show that describing interval-valued wind directions by their centre and
range provides
an alternative approach to the toroidal score-driven method of \cite{mohan2026interval}.

\section{Discussion and conclusion}\label{sec:conclusion}
This paper proposed a centre–-range score-driven approach for interval-valued circular time series. Each interval is described by its circular centre and its range, so each observation lies on a cylinder. Three distributions are considered for the range component: Kumaraswamy, beta, and truncated Weibull. This model is proposed as an alternative to the IN-DCS model, where the lower and upper bounds are jointly modelled in a toroidal, score-driven framework. To our knowledge, this is the first ever centre-range model proposed for circular interval-valued observations. The simulation study shows that the estimation approach is accurate. In both real-world applications, the CR-TWVM, CR-BvM and CR-KvM models provide a better fit as compared to the IN-DCS model. The two representations of circular-valued interval observations are therefore complementary and should be applied accordingly to the shape of the observed intervals in practice. This has also been explored in the synthetic data analysis section, where synthetic data was generated to analyse scenarios for comparing the performance of the IN-DCS and the proposed approaches in this paper. For future work, score-driven approaches can be explored for joint models capable of modelling interval-valued wind direction and wind speed, which would be of applicable use in wind energy applications.

\bibliographystyle{apalike}
\bibliography{references}
\end{document}